\documentclass[11pt,oneside,fleqn]{article}

\usepackage[ansinew]{inputenc}
\usepackage{hyperref}
\usepackage{amsmath,amssymb,amsthm}
\usepackage{graphicx}
\usepackage{multirow}
\usepackage{cite}

\allowdisplaybreaks

\hypersetup{colorlinks, linkcolor=blue, citecolor=blue, urlcolor=blue}
\renewcommand{\arraystretch}{1.2}

\makeatletter
\long\def\@makecaption#1#2{%
  \vskip\abovecaptionskip\footnotesize
  \sbox\@tempboxa{#1. #2}%
  \ifdim \wd\@tempboxa >\hsize
    #1. #2\par
  \else
    \global \@minipagefalse
    \hb@xt@\hsize{\hfil\box\@tempboxa\hfil}%
  \fi
  \vskip\belowcaptionskip}
\makeatother

\newcommand{\pdl}[2]{\frac{\partial #1}{\partial #2}}

\newcommand{\ddd}{\mathrm{d}}
\newcommand{\p}{\partial}
\newcommand{\const}{\mathop{\rm const}\nolimits}

\newcommand{\todo}[1][\null]{\ensuremath{\clubsuit}}

\newcommand{\noprint}[1]{}

\newtheorem{theorem}{Theorem}

{\theoremstyle{definition}
\newtheorem{definition}{Definition}
\newtheorem{example}{Example}
\newtheorem{remark}{Remark}
\newtheorem*{remark*}{Remark}
}

\newcommand{\checked}[1][\null]{\ensuremath{\boldsymbol{\surd}}}

\newcommand{\DD}{\mathrm{D}}

\newcommand{\ve}{\varepsilon}

\begin{document}

\par\noindent {\LARGE\bf
Conservative parameterization schemes
\par}

{\vspace{4mm}\par\noindent {Alexander Bihlo$^\dag$ and George Bluman$^\ddag$
} \par\vspace{2mm}\par}

{\vspace{2mm}\par\noindent {\it
$^{\dag}$~Centre de recherches math\'{e}matiques, Universit\'{e} de Montr\'{e}al, C.P.\ 6128, succ.\ Centre-ville,\\
$\phantom{^\dag}$~Montr\'{e}al (QC) H3C 3J7, Canada
}}

{\vspace{2mm}\par\noindent {\it
$^{\ddag}$~Department of Mathematics, University of British Columbia\\
$\phantom{^\dag}$~Vancouver (BC) V6T 1Z2, Canada
}}

{\vspace{2mm}\par\noindent {\it
$\phantom{^\dag}$~\textup{E-mail}:
$^{\dag}$bihlo@crm.umontreal.ca,
$^{\ddag}$bluman@math.ubc.ca
}\par}

\vspace{4mm}\par\noindent\hspace*{8mm}\parbox{140mm}{\small
Parameterization (closure) schemes in numerical weather and climate prediction models account for the effects of physical processes that cannot be resolved explicitly by these models. Methods for finding physical parameterization schemes that preserve conservation laws of systems of differential equations are introduced. These methods rest on the possibility to regard the problem of finding conservative parameterization schemes as a conservation law classification problem for classes of differential equations. The relevant classification problems can be solved using the direct or inverse classification procedures. In the direct approach, one starts with a general functional form of the parameterization scheme. Specific forms are then found so that corresponding closed equations admit conservation laws. In the inverse approach, one seeks parameterization schemes that preserve one or more pre-selected conservation laws of the initial model. The physical interpretation of both classification approaches is discussed. Special attention is paid to the problem of finding parameterization schemes that preserve both conservation laws and symmetries. All methods are illustrated by finding conservative and conservative invariant parameterization schemes for systems of one-dimensional shallow-water equations.
}\par\vspace{2mm}

\section{Introduction}

The problem of replacing the continuous governing equations of the atmosphere--ocean system by a discrete approximation is that in general no numerical scheme is capable of preserving all the geometrical features that the initial system of differential equations possesses. Among these features are symmetries and conservation laws. The lack of a numerical scheme in preserving fundamental properties of the model has far-reaching consequences on the practical utility of the computed results. Simulating the earth system is a relevant but highly complex task and it involves an intricate interaction of theoretical insight, data handling and numerical modeling. Introducing errors in any of these tasks can lead to severe drifts of the forecasted state towards wrong attractors and thus to misleading weather and climate predictions. To remedy this challenge for the discretization part of the equations, several structure-preserving numerical integrators were developed~\cite{fran04Ay,gass08Ay,somm09Ay,somm10Ay} that might eventually replace the standard integration schemes.

Less research has been carried out so far on the slightly different but related problem of finding structure-preserving closure models or \emph{parameterization schemes} for the subgrid-scale terms that inevitably arise when discretizing a nonlinear system of partial differential equations owing to the limited resolution one has to employ when integrating a numerical model. Attention to the importance of this issue was brought up in~\cite{ober97Ay}, where the task of finding invariant subgrid-scale closure schemes for the filtered Navier--Stokes equations was investigated. These ideas were recently picked up in~\cite{bihl11Fy,popo10Cy} with the aim of formulating general algorithms for finding local parameterization schemes with prescribed invariance characteristics. These methods rely on the property that any generic parameterization ansatz, when introduced into an averaged system of differential equations turns this system into a class of closed differential equations. There exist powerful methods from the field of group analysis of differential equations~\cite{basa01Ay,bihl11Dy,blum89Ay,blum10Ay,blum74Ay,card11Ay,olve86Ay,ovsi82Ay,popo10Ay,popo08Ay} that can be used to study the symmetry properties of such classes of differential equations in an algorithmic way. By a proper interpretation of these methods, one can turn them into effective tools for the construction and study of local parameterization schemes preserving invariance properties.

The fact that the parameterization problem in general can be regarded as the study of the properties of classes of differential equations is also the key for the analog problem of finding parameterization schemes preserving conservation laws. Paralleling the task of finding parameterization schemes that preserve symmetry properties, this problem has important physical applications. Preserving physical conservation laws in the parameterization process is a natural requirement as there exist various processes that preserve e.g.\ energy or mass, but that cannot be resolved in a particular numerical model and thus have to be modeled in a simplified manner. When constructing closure models for such processes it is natural to require the closed system of equations to still preserve energy or mass as otherwise the physical consistency of the parameterization scheme is necessarily violated. Developing algorithmic methods that allow one to construct such parameterization schemes is thus a worthwhile endeavor.

A further point in favor of an extension of the tool kit of geometric closure schemes to include conservative parameterizations is related to the fact that symmetries and conservation laws of a system of differential equations $\mathcal L$ have different relations to the solutions of $\mathcal L$. By definition, a symmetry of $\mathcal L$ is a property of $\mathcal L$ itself without regard to posed initial and boundary conditions whereas a conservation law is a property holding for every solution of $\mathcal L$ irregardless of posed initial and boundary conditions. In particular, most of the nontrivial symmetries of equations in hydrodynamics are broken once classical boundary conditions are imposed (e.g.\ rigid walls or even periodic domains). In contrast, the validity of a conservation law holds in the presence of all kinds of initial and boundary conditions.

Hence it is clear that it is physically relevant to study the relation between conservation laws, classes of differential equations and the parameterization problem. Shedding light on this relation is the purpose of the present paper. The main result of our study is a step towards the first systematic description of general methods for the construction of parameterization schemes that preserve or possess particular conservation laws.

This paper in organized as follows. In Section~\ref{sec:TheoryConservativeParameterizationSchemes}, the theory of conservative parameterization schemes is developed. Here, necessary terminology on conservation laws and group classification is introduced to demonstrate that the problem of finding conservative parameterization schemes can be regarded as a classification problem of conservation laws in classes of differential equations. This classification problem can be solved using both direct and inverse methods. These methods and the additional requirement on conservative parameterization schemes to be invariant with respect to a nontrivial symmetry can successfully be used to narrow down the vast possibility one generally has when constructing local subgrid-scale closure schemes for averaged or filtered differential equations. As an example, conservative and invariant conservative parameterization schemes are constructed for the system of one-dimensional shallow-water equations in Section~\ref{sec:ConservativeParameterizationsShallowWater}. In Section~\ref{sec:ConclusionConservativeParameterizations}, the results of the paper are summarized and suggestions are given for further research directions.

\section{Conservative and invariant parameterization schemes}\label{sec:TheoryConservativeParameterizationSchemes}

In this section we introduce some necessary terminology on symmetries and conservation laws, which are essential to formulate the theory of invariant and conservative parameterization schemes in a proper way. The exposition of the background material follows~\cite{anco02Ay,anco02By,blum10Ay,olve86Ay,popo10Cy,popo10Ay,popo08Ay}, to which we refer for a more thorough discussion of the underlying notions, methods and theoretical concepts.

\subsection{General notions and statement of the problem}

Let there be given a system of differential equations denoted by $\mathcal L$, which consists of $L$ equations of the form $\Delta_l(x,u^{(n)})=0$, $l=1,\dots L$, where $x=(x^1,\dots,x^p)$ are the independent variables, $u=(u^1,\dots,u^q)$ are the dependent variables and $u^{(n)}$ denote all the derivatives of $u$ with respect to $x$ up to order $n$, with $u$ being included as the zeroth order derivative.

\begin{definition}
 A \textit{local conservation law} of the system $\mathcal L$ is a divergence expression
 \begin{equation}\label{eq:ConservationLaw}
  \DD_j\Phi^j|_{\mathcal L}=(\DD_1\Phi^1+\cdots+\DD_p\Phi^p)|_{\mathcal L}=0,
 \end{equation}
 which holds on the solution space of the system $\mathcal L$ (denoted by $|_{\mathcal L}$). The $p$-tuple of differential functions $\Phi=(\Phi^1(x,u^{(m)}),\dots,\Phi^p(x,u^{(m)}))$, $m\in\mathbb{N}_0$, is a \textit{conserved vector} of the associated conservation law.
\end{definition}

Here and in the following, $\DD_i$ is the operator of total differentiation with respect to $x^i$, defined by $\DD_i=\p_{x^i}+u^\alpha_{J,i}\p_{u^\alpha_J}$, where $u^\alpha_J=\partial^{|J|}u^\alpha/\partial (x^1)^{j_1}\cdots\partial (x^p)^{j_p}$, $u^\alpha_{J,i}=\partial u^\alpha_J/\partial x^i$, $J=(j_1,\dots,j_p)$ is a multi-index, $j_i\in\mathbb{N}_0$ and $|J|=j_1+\dots+j_p$. The summation convention over repeated indices is understood.

\begin{definition}
 A local conservation law of the system $\mathcal L$ is \textit{trivial} if the components of the conserved vector $\Phi$ are of the form $\Phi^j=M^j(x,u^{(m)})+H^j(x,u^{(m)})$, where the differential function $M^j$ vanishes on the solution space of the system $\mathcal L$ and the differential function $H^j$ is a null divergence, i.e.\ it satisfies $\DD_jH^j=0$ identically.
\end{definition}

As trivial conservation laws satisfy the divergence condition unrestricted by the system of differential equations $\mathcal L$, they contain no relevant information about the system $\mathcal L$. Thus, only nontrivial conservation laws are of interest below.

\begin{definition}\label{def:EquivalentConservationLaws}
 Two conserved vectors $\Phi$ and $\Phi'$ represent the same conservation law (i.e.\ are \textit{equivalent}) if their difference $\Phi-\Phi'$ is a conserved vector associated with a trivial conservation law.
\end{definition}

The above definition implies that conservation laws can only be found up to composition with trivial conservation laws, i.e.\ there is not a single canonical representation of one and the same conservation law. Thus, formally the space of conservation laws can be defined as the set of elements from the factor space of the set of all conserved vectors with respect to the subset of trivial conserved vectors. See~\cite{popo08Ay} for more details.

Conservation laws are conveniently found using the \emph{multiplier approach}. This method rests on an equivalent recasting of the definition of a conservation law~\eqref{eq:ConservationLaw} in the form
\begin{equation}\label{eq:ConservationLawMultiplier}
 \DD_j\Phi^j(x,U^{(m)}) = \Lambda^l(x,U^{(r)})\Delta_l(x,U^{(n)}),
\end{equation}
where the differential functions $\Lambda =(\Lambda^1(x,U^{(r)}),\dots,\Lambda^L(x,U^{(r)}))$, $r\in\mathbb{N}_0$ are \textit{conservation law} (CL) \textit{multipliers}, also called the \emph{characteristics of the conserved vector} $\Phi$. Note that expression~\eqref{eq:ConservationLawMultiplier} holds for arbitrary functions $U(x)$. It is then obvious that for solutions $U(x)=u(x)$ of the system $\mathcal L$ the right-hand side of the above expression vanishes and thus~\eqref{eq:ConservationLawMultiplier} reduces to the definition of a conservation law given above, provided that the characteristic $\Lambda$ is non-singular on the solution manifold of $\mathcal L$.

Expression~\eqref{eq:ConservationLawMultiplier} can be converted into a system of determining equations for the multipliers~$\Lambda$. This is facilitated by means of the Euler operator.

\begin{definition}
 The \emph{Euler operator} with respect to the dependent variable $U^i$ is the differential operator given by
 \begin{equation}\label{eq:EulerOperators}
  \mathsf{E}_i = \partial_{U^i}-\DD_{j_1}\partial_{U^i_{j_1}}+\DD_{j_1}\DD_{j_2}\partial_{U^i_{j_1j_2}}-\cdots= (-\DD)^J\partial_{U^i_J},
 \end{equation}
 where $(-\DD)^J=(-\DD_1)^{j_1}\dots(-\DD_p)^{j_p}$.
\end{definition}

The importance of Euler operators in the study of local conservation laws lies in the property that they annihilate any divergence expression $\DD_j\Phi^j$. In particular the CL multipliers $\Lambda = (\Lambda^1(x,U^{(r)}),\dots,\Lambda^L(x,U^{(r)}))$, $r\in\mathbb{N}_0$, yield a CL of $\mathcal L$ if and only of
\begin{equation}\label{eq:DeterminingEquationsMultipliers}
 \mathsf{E}_i(\Lambda^l\Delta_l)\equiv0,\quad i=1,\dots,q.
\end{equation}
Equation~\eqref{eq:DeterminingEquationsMultipliers} can be split with respect to $\Delta_l$ and its differential consequences. This yields an over-determined linear system of partial differential equations, which serve as the determining equations for the local CL multipliers of the system~$\mathcal L$. Once these multipliers are found, the associated conserved vectors $\Phi$ can be constructed using e.g.\ the \emph{direct construction method}~\cite{anco02Ay,anco02By,blum10Ay}.

We now move on to the precise statement of the \emph{parameterization problem}. Let there be given a system of differential equations $\mathcal L$, $\Delta_l(x,u^{(n)})=0$, $l=1,\dots,L$ and let there be defined a filtering operation
\begin{equation}\label{eq:GeneralFormOfAveragingOperator}
 \mathcal P(u^i)=\bar u^i(x) = \int_\Omega u^i(y)G(x,y)\,\ddd y,
\end{equation}
where $\ddd y = \ddd y^1\cdots\ddd y^p$ and $\Omega=\int\ddd y$. Eq.~\eqref{eq:GeneralFormOfAveragingOperator} is the convolution of the variable $u^i$ with the filter kernel $G=G(x,y)$. The filter kernel $G(x,y)$ satisfies
\[
 \int_\Omega G(x,y)\,\ddd y=1,
\]
see~\cite{germ92Ay,ober97Ay,saga05Ay}. This averaging operation can be used to decompose the instantaneous dependent variables $u$ according to
\[
 u=\bar u+u'.
\]
The average $\bar u$ is referred to as the resolved or grid-scale part of the dynamics, while $u'$ includes the unresolved subgrid-scale fraction of $u$. As we do not mix different averaging methodologies in one and the same physical problem, we subsequently denote by a bar any mean value of $u$, irrespectively of what averaging operator is used in the concrete problem of interest.

\begin{example}\label{ex:ReynoldsAveraging}
In the classical \textit{Reynolds averaging} one uses the time average of $u$, which is defined~by
\[
 \mathcal P_{R}(u^i)=\bar u^i(x^*) = \lim_{T\to\infty}\frac{1}{T}\int_{t_0}^{t_0+T}u^i(t,x^*)\,\ddd t,
\]
where $t_0$ denotes the initial time one starts to average. This time average follows from~\eqref{eq:GeneralFormOfAveragingOperator} upon setting $t=x^1$, $x^*=(x^2,\dots,x^p)$ and by factorizing
\[
 G(x,y)=G(t-y^1)\prod_{i=2}^pG_i(x^i-y^i)=\frac{H_T}{T}\prod_{i=2}^p\delta(x^i-x^j),
\]
where $H_T$ is the step function over the interval $T$ and $\delta$ is the delta distribution. The time averaging is a Reynolds operator, i.e.\ it satisfies $\overline{\bar u_iu_j}=\bar u_i\bar u_j$. Owing to this property, in the splitting $u=\bar u+u'$ one has $\overline{u'}=0$ since $\bar{\bar u}=\bar u$. The average over a product $u^iu^j$ thus gives $\overline{u^iu^j}=\bar u^i\bar u^j+\overline{u^{i'}u^{j'}}$, which is the classical Reynolds decomposition that introduces the Reynolds stresses $\overline{u^{i'}u^{j'}}$ into the averaged Navier--Stokes equations. In practical computation a finite $T<\infty$ has to be chosen and then $\bar u=\bar u(t,x^*)$, i.e.\ the mean value is still time-dependent.
\end{example}

\begin{example}\label{ex:Filtering}
In \textit{large--eddy simulation} of turbulence, the classical Reynolds averaging as introduced in Example~\ref{ex:ReynoldsAveraging} is replaced by a spatial filtering approach defined by
\[
 \mathcal P_{LES}(u^i)=\bar u^i(t,x^*) = \int_\Omega u^i(t,y^*) G(x,y)\ddd y,
\]
for which the filter kernel in~\eqref{eq:GeneralFormOfAveragingOperator} is decomposed according to
\[
  G(x,y)=G(t-y^1)\prod_{i=2}^pG_i(x^i-y^i)=\delta(t-y^1)\prod_{i=2}^pG_i(x^i-y^i).
\]
The filters defined in this way are generally not Reynolds operators, i.e.\ now $\overline{u'}\ne0$ as $\bar{\bar u}\ne\bar u$ and thus filtering over products $u^iu^j$ produces additional terms of forms not present in the Reynolds averaging approach, i.e.\ $\overline{u^iu^j}=\overline{\bar u^i\bar u^j}+\overline{u^{i'}\bar u^j}+\overline{\bar u^iu^{j'}}+\overline{u^{i'}u^{j'}}$.
\end{example}

With the aid of a particular averaging operator~\eqref{eq:GeneralFormOfAveragingOperator} the system $\mathcal L$ is converted into a system for the resolved part $\bar u$, which can be determined by measurements or in the course of a numerical simulation of the system~$\mathcal L$. This is done by introducing the splitting $u=\bar u+u'$ into the system $\mathcal L$ followed by an application of a specific filtering~\eqref{eq:GeneralFormOfAveragingOperator}, which leads to the averaged system of differential equations $\mathcal{\bar L}$ given by
\begin{equation}\label{eq:GeneralFormOfAveragedEquation}
 \bar \Delta_l (x,\bar u^{(n)},w)=0,\quad l=1,\dots,L.
\end{equation}
In this expression the $k$-tuple $w=(w^1,\dots,w^k)$ includes all those terms that arise in the course of the averaging or filtering and cannot be determined from the knowledge of the mean or filtered values $\bar u^{(n)}$. For the Reynolds averaging introduced in Example~\ref{ex:ReynoldsAveraging}, these are terms like $\overline{u^{i'}u^{j'}}$ (or higher order products as well as their derivatives), while in the case of the spatial filtering of Example~\ref{ex:Filtering}, $w$ would additionally include terms of the form $\overline{u^{i'}\bar u^j}$, etc. This is essentially the closure problem, i.e.\ Eqs.~\eqref{eq:GeneralFormOfAveragedEquation} include more unknown than known quantities. Thus, as they stand Eqs.~\eqref{eq:GeneralFormOfAveragedEquation} cannot be used for a numerical integration unless one expresses the additional unknowns~$w$ in terms of certain known expressions.

\begin{definition}
A \textit{local} parameterization or subgrid-scale \textit{closure} model assumes a functional relation between the unknown subgrid-scale terms $w$ and the mean values $\bar u^{(r)}$, $r\in\mathbb{N}_0$ i.e.\
\begin{equation}\label{eq:GeneralFormOfParameterizationScheme}
 w^i=f^i(x,\bar u^{(r)}), \quad i=1,\dots,k,
\end{equation}
for certain \textit{parameterization functions} $f=(f^1,\dots,f^k)$.
\end{definition}

Introducing a local parameterization scheme~\eqref{eq:GeneralFormOfParameterizationScheme} into system $\mathcal {\bar L}$~\eqref{eq:GeneralFormOfAveragedEquation} leads to a closed system of differential equations for the mean values $\bar u$. The inherent problem of this construction is that in most cases of interest the information contained in $\bar u$ and its derivatives is not sufficient to determine the entire subgrid-scale structure contained in $w$. The \textit{art} of constructing physical parameterization schemes is to determine the parameterization functions $f$ in such a manner that the assumption~\eqref{eq:GeneralFormOfParameterizationScheme} will allow one to find $\bar u$ from the closed system $\mathcal{\bar L}$ with sufficient accuracy.

Finding suitable parameterization functions $f$ that lead to realistic results for $\bar u$ can be rather tedious. One general methodology to restrict the vast number of possible forms for the parameterization functions is to choose them in such a manner that the resulting closed system of differential equations preserves certain nontrivial geometric properties such as conservation laws and/or symmetries. This motivates the following definition.

\begin{definition}
 A local parameterization scheme is called \textit{conservative} provided that the closed class of differential equations $\mathcal{\bar L}$ preserves certain nontrivial local conservation laws. A local parameterization scheme is called \textit{invariant} provided that the closed class of differential equations $\mathcal{\bar L}$ preserves a nontrivial point symmetry group $G$.
\end{definition}

Conservative parameterization schemes can be found using techniques analogous to those for the \emph{group classification} of classes of differential equations. A \emph{class of differential equations} $\mathcal L|_{\mathcal S}$ is a system of differential equations of the form $\Delta_l(x,u^{(n)},\theta(x,u^{(n)}))=0$, $l=1,\dots,L$, which is parameterized by a $k$-tuple $\theta=(\theta^1,\dots,\theta^k)$ of differential functions that satisfy a system of $K\in\mathbb{N}_0$ auxiliary differential equations of the form $S_j(x,u^{(n)},\theta^{(m)}(x,u^{(n)}))=0$, $j=1,\dots,K$, the solution set of which is denoted by $\mathcal S$. The system of auxiliary equations in part specifies the properties of the class and it is regarded as a system for $\theta$, i.e.\ $x$ and $u^{(n)}$ play the role of independent variables. To complete the description of the class $\mathcal L|_{\mathcal S}$ one usually takes into account a constitutive inequality, $\Sigma(x,u^{(n)},\theta^{(m)}(x,u^{(n)}))\ne0$, which guarantees that all equations from the class share some joint properties (e.g.\ a particular derivative does not vanish, all equations of $\mathcal L|_{\mathcal S}$ are linear or nonlinear, etc.).

It is the purpose of conservation law classification to systematically investigate the CLs of a class of differential equations. By substituting the general closure scheme~\eqref{eq:GeneralFormOfParameterizationScheme} into the averaged system~$\mathcal {\bar L}$~\eqref{eq:GeneralFormOfAveragedEquation} one obtains a \emph{class of closed differential equations} for $\bar u$,
\begin{equation}\label{eq:GeneralFormOfAveragedEquationClosed}
 \bar \Delta_l (x,\bar u^{(n)},f(x,\bar u^{(r)}))=0,\quad l=1,\dots,L,
\end{equation}
in which the parameterization functions $f$ play the role of the arbitrary elements $\theta$. In the following subsections we will introduce and discuss methods that allow one to specify the parameterization functions~$f$ in such a manner that the system~\eqref{eq:GeneralFormOfAveragedEquationClosed} has certain nontrivial local CLs. The corresponding classification methods for finding parameterization schemes that possess nontrivial maximal Lie invariance groups where introduced in~\cite{blum74Ay}, see also~\cite{popo10Cy}. Combinations of invariant and conservative parameterization schemes are also possible and will be discussed at the end of this section.

\begin{remark}\label{rem:OnExtensionsOfDifferentialEquations}
 A problem related to the search for parameterization or closure schemes for the subgrid-scale terms arising in averaged differential equations is to search for \emph{extensions} of differential equations that preserve some of the geometric features of the original differential equations $\mathcal L$. Physically, such extensions could be e.g.\ adding dissipation terms to a non-dissipative model or source or sink terms to transport equations. Depending on the nature of the included process, the addition of such extra terms may alter the structure of the initial system of differential equation but might still retain some of the geometric features of the original model. A possible research question is thus to construct extra terms for the system $\mathcal L\colon \Delta_l(x,u^{(n)})=0$, $l=1,\dots L$ in such a manner as to preserve certain CLs and/or symmetries of $\mathcal L$. Mathematically, this is done by investigating systems of the form $\Delta_l(x,u^{(n)})=g_l(x,u^{(r)})$, for a certain $L$-tuple $g=(g^1(x,u^{(r)}),\dots g^L(x,u^{(r)}))$ of extensions. It is obvious that this system can be brought into the form~\eqref{eq:GeneralFormOfAveragedEquationClosed} if the functions $f$ are interpreted as the additional terms that extend the initial system $\mathcal L$ and no averaging operation is involved, i.e.\ $\bar u^{(n)}=u^{(n)}$. Consequently, the same methods as introduced below for solving the parameterization problem for system~\eqref{eq:GeneralFormOfAveragedEquationClosed} can be used to tackle this kind of problem.
\end{remark}

\subsection{Conservative parameterizations via direct classification}

In order to discuss the method for finding conservative parameterization schemes, the following definitions are useful.

\begin{definition}
 An \textit{equivalence transformation} $\varphi$ from the class $\mathcal L|_{\mathcal S}$ is a point transformation on the space $(x,u^{(n)},\theta)$, which is projectable on the spaces of $(x,u^{(n')})$, $0\le n'\le n$, such that $\forall \theta\in\mathcal S\colon \theta'=\varphi\theta\in\mathcal S$ and the restriction of $\varphi$ to the space of $(x,u^{(n)})$, denoted by $\varphi|_{(x,u^{(n)})}$, is a point transformation from $\mathcal L_\theta$ to $\mathcal L_{\theta'}$. Here, $\mathcal L_{\theta}$ and $\mathcal L_{\theta'}$ are equations from the class $\mathcal L|_{\mathcal S}$.
\end{definition}

Thus, equivalence transformations are point transformations that map one system of differential equations from a given class $\mathcal L|_{\mathcal S}$ to another system of differential equations from the same class. The collection of all equivalence transformations forms a group, which is called the \textit{equivalence group} $G^\sim$.

\begin{definition}\label{def:EquivalenceConservationLawsClasses}
 Let there be given two systems of differential equations from the class $\mathcal L|_{\mathcal S}$, denoted by $\mathcal L_\theta$ and $\mathcal L_{\theta'}$, which have CLs with conserved vectors $\Phi$ and $\Phi'$, respectively. The pairs $(\mathcal L_\theta,\Phi)$ and $(\mathcal L_{\theta'},\Phi')$ are \textit{equivalent} with respect to the equivalence group $G^\sim$ if there exists a point transformation $\varphi\in G^\sim$ that transforms the system $\mathcal L_\theta$ to the system $\mathcal L_{\theta'}$ and which transforms the conserved vector $\Phi$ in such a manner that $\tilde \Phi=\varphi(x,u^{(r)},\Phi)$ and $\Phi'$ are equivalent conservation laws, see Definition~\ref{def:EquivalentConservationLaws}.
\end{definition}

In this definition, the action of a point transformation $\varphi \in G^\sim$ on a conserved vector $\Phi$ has the explicit form
\[
 \tilde \Phi^i(\tilde x,\tilde u^{(r)})=\frac{\DD_{x^j}\tilde x^i}{|\DD_{x}\tilde x|}\Phi^j(x,u^{(r)}),\quad  i=1,\dots,p,
\]
where $|\DD_{x}\tilde x|$ is the determinant of the matrix $(\DD_{x^j}\tilde x^i)$. See~\cite{blum06Ay,blum10Ay,popo08Ay} for more details.

The direct classification procedure for finding parameterization schemes with prescribed CLs can be formulated in the following way. For a given fixed general form of the parameterization functions $f$, determine those CLs that hold for any equation from the class~\eqref{eq:GeneralFormOfAveragedEquationClosed} (i.e.\ for all admissible forms of $f$) and find all the inequivalent equations from that class that have additional CLs.

To make the classification problem tractable, one first chooses the general form of parameterization functions $f$ one aims to study, i.e.\ one determines which variables $x$ and $u^{(r)}$ the functions $f$ should depend on. \emph{This choice is physically motivated}. Once the general form of $f$ (hence the system of auxiliary equations $\mathcal S$) is fixed, one can solve the classification problem taking into account the equivalence of CLs as embodied in Definition~\ref{def:EquivalenceConservationLawsClasses}. This means that one determines the equivalence group of the general class of closed differential equations of interest and then solves the determining equations~\eqref{eq:DeterminingEquationsMultipliers} for CL multipliers. One seeks those values of $f$ (up to equivalence) for which the determining equations for CL multipliers yield additional multipliers beyond those for generic parameterization functions. An example for this procedure is given in Section~\ref{sec:ShallowWaterDirectMethod}.

The direct group classification method yields a list of inequivalent equations $\mathcal {\bar L}_f$ from the class $\mathcal {\bar L}|_{\mathcal S}$ that possess inequivalent nontrivial local CLs. Using this list of all possible conservative parameterization schemes from the predefined class, one can then test the different schemes obtained and select the most appropriate one as a candidate closure scheme for the process of interest that needs to be parameterized.

Physically, the method of finding conservative parameterization schemes using the direct classification approach might be most appropriate in the case when one seeks to represent processes that are not already included in the dynamics resulting from the system~$\mathcal{L}$. The reason for this is that by means of the direct classification method one might obtain closed differential equations that have CLs not possessed by the original system~$\mathcal{L}$.

\subsection{Conservative parameterizations via inverse classification}\label{sec:TheoryInverseClassification}

A different ideology for finding parameterization schemes is the following. Assume that the original system of differential equations $\mathcal L$ has a certain number of nontrivial local CLs. The averaging of a differential equation certainly disturbs the geometric structure of the equation but it might nevertheless be desirable that the averaged system share some CLs of the original system of equations. An example for this is a process that conserves energy but needs to be parameterized in a given system of differential equations. For the sake of physical consistency, the closed differential equations should conserve energy and thus only parameterization schemes that are compatible with energy conservation can be considered.

This discussion is related to what is called the \emph{inverse classification problem} and in the framework of a conservative parameterization scheme, it can be formulated in the following way. Let there be given an initial system of differential equations $\mathcal L$. One first determines CLs holding for the original system of differential equations $\mathcal L\colon \Delta_l(x,u^{(n)})=0$, $l=1,\dots L$ through the CL multiplier approach. Depending on the complexity of the problem of interest, one might not be able to obtain an exhaustive description of all CLs but rather restricts oneself to CLs associated with characteristics $\Lambda(x,u^{(r)})$ for some fixed (often low-dimensional) $r$. Among the CLs of $\mathcal L$, one selects, using physical reasoning, the associated multipliers of those CLs that one aims to preserve in the course of the parameterization process. As in the case of the direct classification method, one next averages the system $\mathcal L$ and determines the general functional form of the parameterization functions $f$ to be used in the class of parameterization schemes~\eqref{eq:GeneralFormOfParameterizationScheme}. The final step is to plug the class of averaged closed differential equations~\eqref{eq:GeneralFormOfAveragedEquationClosed} into the determining equations~\eqref{eq:DeterminingEquationsMultipliers} for CL multipliers. Since the multipliers that the resulting equations from the class~\eqref{eq:GeneralFormOfAveragedEquationClosed} should admit are fixed, the determining equations for local CL multipliers are thus converted into a system of determining equations for the parameterization functions $f$. Solving this system leads to all equations from the class $\mathcal {\bar L}|_{\mathcal S}$ that have the same CL multipliers $\Lambda(x,\bar u^{(r)})$ as the original system $\mathcal L$, with $u^{(r)}$ being replaced by the corresponding averaged values $\bar u^{(r)}$.

A nontrivial question in this construction procedure is to determine in advance whether at least some systems from the class $\mathcal {\bar L}|_{\mathcal S}$ selected has the CLs associated with the chosen multipliers $\Lambda(x,u^{(r)})$ stemming from the original system $\mathcal L$, i.e.\ whether the determining equations~\eqref{eq:DeterminingEquationsMultipliers} yield any nontrivial solutions. A natural strategy to overcome this problem of possibly triviality of the solution of~\eqref{eq:DeterminingEquationsMultipliers} is to (i) either assume that the class of closed equations $\mathcal {\bar L}|_{\mathcal S}$ is rather wide (i.e.\ that the function $f$ depends on a large number of variables from $\bar u^{(r)}$) or (ii) to only require a suitable small set of CLs being preserved by the parameterization scheme. From the physical point of view, the first strategy should be the method of choice.

Although this possible triviality of the solution of the determining equations for the parameterization functions and the associated failure in finding nontrivial conservative parameterization schemes seems undesirable, it nevertheless includes important physical evidence. It simply means that for the parameterization ansatz selected, no element of the class of closed equations can satisfy the requirement of retaining the desired CLs and thus might indicate that the initial parameterization ansatz was flawed.

In contrast to the direct classification method, which might lead to conservative parameterization schemes that possess CLs which do not hold for the original system~$\mathcal L$, conservative parameterization schemes derived using the inverse classification procedure by construction yield no additional CLs for the resulting closed system of differential equations. The inverse classification strategy might thus be best suited for processes already included in the full dynamics of the original system~$\mathcal L$, but that cannot be explicitly resolved because of e.g.\ computational limitations.

\begin{remark}\label{rem:OnExistenceOfSolutionConservativeSchemesInverse}
 The existence of at least the trivial solution of Eqs.~\eqref{eq:DeterminingEquationsMultipliers} in the case when the CL multipliers are fixed and possible forms of $f$ are sought, i.e.\ that the system of determining equations is compatible is guaranteed by the fact that~\eqref{eq:GeneralFormOfAveragedEquationClosed} can be rewritten in the form
 \[
   \Delta_l (x,\bar u^{(n)})=g_l(x,\bar u^{(r)}),\quad l=1,\dots,L,
 \]
 where the left-hand side is the same as in the original system $\mathcal L$ provided that $\bar u^{(n)}$ is used in place of $u^{(n)}$ and the right-hand side is a functional combination of the parameterization functions $f$. Thus if $g=0$ (and hence $f=0$), the above equation admits the same CL multipliers $\Lambda$ as for the system $\mathcal L$ in which $\bar u^{(n)}$ is used instead of $u^{(n)}$.
\end{remark}

\subsection{Conservative and invariant parameterizations}

Methods for finding parameterization schemes with symmetry properties using methods from the group analysis of differential equations were introduced in~\cite{bihl11Fy,blum74Ay,popo10Cy}. There is neither a practical nor a theoretical objection against a parameterization scheme preserving both invariance \textit{and} CL properties. Indeed, the compatibility of these two concepts was explicitly demonstrated by constructing CL and invariance preserving hyperdiffusion schemes for the two-dimensional barotropic vorticity equation on the beta-plane~\cite{bihl11Fy}. We now outline how to systematically construct \textit{invariance and CL preserving parameterization schemes}.

As in the case of conservative parameterization schemes, two main methods are applicable to determine parameterization schemes with symmetry properties. These methods are straightforward applications of the group analysis of differential equations and the required techniques are based on the \textit{direct} and the \textit{inverse} approach to the symmetry classification problem, respectively. The key to the construction of conservative invariant parameterization schemes is that the closure models resulting from the conservative parameterization procedure as outlined in the previous two subsections are usually still classes of differential equations. These classes are generally narrower than the initial class given by Eqs.~\eqref{eq:GeneralFormOfAveragedEquationClosed} but nevertheless include arbitrary constants or parameter functions with respect to which the usual symmetry classification problem can be carried out. We only outline the main ideas of this construction below as a more detailed exposition of the methods available in the field would require a substantial extension of the text. Moreover, the symmetry analysis and group classification of differential equations is a well-investigated subject. See~\cite{anco02Ay,anco02By,anco08Ay,basa01Ay,bihl11Dy,blum89Ay,blum10Ay,blum74Ay,card11Ay,olve86Ay,olve09Ay,ovsi82Ay,popo10Ay,popo08Ay} and references therein.

Special attention will be also paid to models that are derivable from \textit{variational principles}, i.e.\ which are \textit{Euler--Lagrange equations}. For such systems there is a close connection between symmetries and CLs that can be utilized to construct invariant conservative parameterization schemes. In particular, in this situation, all CL multipliers are symmetries but the converse is false~\cite{anco02Ay,anco02By,blum89Ay,blum10Ay,olve86Ay}.

\medskip

\noindent\textbf{Invariant conservative parameterizations via direct symmetry classification.} In the direct symmetry classification approach one starts with a given class of differential equations and determines those symmetries that hold for all equations in the class. These symmetries form the \emph{kernel} of maximal Lie invariance groups of equations in the class. The direct classification problem is solved by finding all equations in the class that have symmetry extensions with respect to the kernel and this investigation is carried out up to the equivalence imposed by the equivalence group~$G^\sim$.

Depending on the complexity of the class of closed differential equations~\eqref{eq:GeneralFormOfAveragedEquationClosed} with conservative properties, different strategies for solving the direct symmetry classification problem can be adopted. For simple classes depending only on a few arbitrary constants or parameter functions with few arguments, the \textit{direct integration of the determining equations} of Lie symmetries up to equivalence using compatibility analysis is the method of choice. This method yields a complete list of inequivalent equations from classes of the form~\eqref{eq:GeneralFormOfAveragedEquationClosed} that are both conservative and have nontrivial symmetry properties. If the structure of the class of~\eqref{eq:GeneralFormOfAveragedEquationClosed} is too complicated for a direct integration of the determining equations of Lie symmetries then the \textit{algebraic method} of group classification can be used. With this method one aims to find symmetry extensions of the kernel that are induced by transformations from the equivalence group of the class under consideration. The algebraic method thus reduces the problem of finding symmetry extensions to the problem of finding inequivalent subgroups of the equivalence group $G^\sim$. Similar to the direct integration of the determining equations, the algebraic method of group classification can lead to the complete solution of the group classification problem, namely for classes of differential equations possessing the normalization property, see~\cite{popo10Ay}. If the given class of equations is not normalized, then the algebraic method will not lead to a complete description of all possible inequivalent symmetry extensions of the kernel. One will still find those symmetry extensions that are induced by the equivalence transformations of the class of equations of interest but there can be other symmetry extensions of the kernel that cannot be found from the classification of the subgroups of the equivalence group. The algebraic method of direct group classification for classes that are not normalized is also known as \textit{preliminary group classification}. A more detailed discussion of the techniques available in the field of direct group classification can be found in~\cite{bihl11Dy,card11Ay,popo10Ay}.

Irrespective of what method is used to (partially) solve the direct symmetry classification problem, all the systems of closed differential equations obtained in the classification procedure have the same CLs but different (inequivalent) maximal Lie invariance groups. The resulting closure schemes then have to be tested numerically to find the most suitable invariant and conservative representation for a given subgrid-scale process.

\medskip

\noindent\textbf{Invariant conservative parameterizations via inverse symmetry classification.} The inverse symmetry classification problem is to find all those equations that have a prescribed symmetry property. See~\cite{blum74Ay} for the first systematic outline of this problem for both ordinary and partial differential equations. Here, rather than starting from a given class of differential equations and describing the invariance properties of equations from this class as done in the direct symmetry classification, in the inverse classification one starts with a given Lie group of transformations and seeks to find the class of equations (up to some order $n$) that is invariant under the selected group. The inverse symmetry classification procedure rests on the following theorem~\cite{blum89Ay,blum10Ay,blum74Ay,olve86Ay,olve09Ay}:

\begin{theorem}
 Let there be given a Lie group of transformations $G$ acting on a manifold $M$. If the $n$th prolongation of $G$ acts regularly on the $n$th order jet space $J^n$ and if there exists a functionally independent system of $n$th order differential invariants $I_1,\dots,I_N$, $I_i=I_i(x,u^{(n)})$, then any $n$th order system of differential equations $\mathcal L$ admitting $G$ as a symmetry group can be rewritten in terms of these differential invariants, i.e.\
 $
  \Delta_l(x,u^{(n)})=\tilde\Delta_l(I_1,\dots,I_N)=0,
 $
 $l=1,\dots,L$.
\end{theorem}

This result is known as the \textit{replacement theorem}~\cite{cheh08Ay,fels99Ay}. It implies the existence of certain nonvanishing multipliers $\Gamma^\kappa_l=\Gamma^\kappa_l(x,u^{(n)})$ that are differential functions such that the following holds for the system of differential equations $\mathcal L\colon \Delta_l(x,u^{(n)})=\tilde\Delta_l(I_1,\dots,I_N)=0$, $l=1,\dots,L$,
\begin{equation}\label{eq:InvariantRepresentationDifferentialEquation}
  \Gamma^\kappa_l(x,u^{(n)})\Delta_\kappa(x,u^{(n)}) = \tilde\Delta_l(I_1(x,u^{(n)}),\dots,I_N(x,u^{(n)})),\quad l =1,\dots,L.
\end{equation}

One can use the replacement theorem to construct parameterization schemes that have specified CL and invariance properties. To this end, one first determines the complete system of $s$th order differential invariants $I_i$ of the maximal Lie invariance group $G$ of the original unaveraged system of differential equations~$\mathcal L$ or of an appropriate subgroup $G^1\subset G$, where $s=\textup{min}(n,r)$, $n$ and~$r$ being the highest order derivatives arising in $\mathcal L$ and the parameterization ansatz~\eqref{eq:GeneralFormOfParameterizationScheme}, respectively. These invariants can be found either using infinitesimal techniques~\cite{blum89Ay,blum10Ay,blum74Ay,olve86Ay,ovsi82Ay} or moving frames~\cite{cheh08Ay,fels98Ay,fels99Ay}.

Secondly, once the invariants of $G$ (or $G^1$) are known one can find the multipliers $\Gamma^\kappa_l$ and thus obtain the invariant representation~\eqref{eq:InvariantRepresentationDifferentialEquation} of the system $\mathcal L$. Now suppose that the parameterization functions $f$ in the class~\eqref{eq:GeneralFormOfAveragedEquationClosed} have been determined in such a manner that the resulting equations from~\eqref{eq:GeneralFormOfAveragedEquationClosed} have the desired CLs. As in Remark~\ref{rem:OnExistenceOfSolutionConservativeSchemesInverse}, one represents~\eqref{eq:GeneralFormOfAveragedEquationClosed} in the solved form $\Delta_l(x,\bar u^{(n)})=g_l(x,\bar u^{(r)})$, $l=1,\dots,L$. These closed equations will be invariant under $G$ (or~$G^1$) for those functions $g$ that satisfy the system of equations
\begin{equation}\label{eq:ConditionForInvariantConservativeSchemes}
 \Gamma^\kappa_l(x,\bar u^{(n)}) g_\kappa(x,\bar u^{(r)})= \tilde g_l(I_1,\dots,I_N),\quad l=1,\dots,L,
\end{equation}
using the same multipliers $\Gamma^\kappa_l$ (replacing $u^{(n)}$ with $\bar u^{(n)}$), for some $N$-tuple of functions $\tilde g$ of the differential invariants $I_i$.

Similar to the discussion in Section~\ref{sec:TheoryInverseClassification}, it is a nontrivial question to determine in advance whether condition~\eqref{eq:ConditionForInvariantConservativeSchemes} can be satisfied for a given set of conservative parameterization functions $f$ and a chosen symmetry group $G$ (or $G^1$). In general, the wider the symmetry group $G$ (or $G^1$) is, the more specific are the forms of the differential invariants $I_i$ and hence the more general the class of conservative parameterization schemes has to be in order to allow jointly for CL and invariance properties. On the other hand, if from the physical point of view it is required that the parameterization of a certain process should have specified invariance and CL properties, then failure in satisfying condition~\eqref{eq:ConditionForInvariantConservativeSchemes} may again point to an inappropriately chosen parameterization ansatz for~$f$ and thus is one more check for the consistency of a physical parameterization scheme.

\medskip

\noindent\textbf{Variational parameterizations for Lagrangian systems.} As mentioned previously, there is a direct close connection between symmetries and CLs for systems of differential equations that can be derived from a variational principle. More precisely, for each one-parameter Lie group of point transformations (or, more generally, one-parameter group of higher order local transformations) that leaves invariant the functional
\begin{equation}\label{eq:ActionFunctional}
 \mathcal S[u] = \int_\Omega L(x,u^{(n)})\ddd x,
\end{equation}
where $\ddd x = \ddd x_1\cdots \ddd x_p$, to within a divergence, there is a local CL of the Euler--Lagrange equations associated with~\eqref{eq:ActionFunctional}. This result is the celebrated \emph{Noether theorem}~\cite{blum89Ay,blum10Ay,olve86Ay}. Practically, the invariance of the functional $\mathcal S[u]$ under point transformations of the form $\tilde x=\tilde x(x,u,\ve)$ and $\tilde u=\tilde u(x,u,\ve)$, with associated infinitesimal generator $Q=\xi^i(x,u)\p_{x^i}+\phi^\alpha(x,u)\p_{u^\alpha}$, can be determined by checking whether the condition
\begin{equation}\label{eq:ConditionOnVariationalSymmetries}
 Q^{(n)}(L)+L\DD_i\xi^i=\DD_iB^i
\end{equation}
is satisfied, where $Q^{(n)}$ denotes the $n$th prolongation of $Q$ and $B=(B^1(x,u^{(m)}),\dots,B^p(x,u^{(m)}))$ is some $p$-tuple of differential functions. In this case, it can be proved~\cite{blum89Ay,blum10Ay,olve86Ay} that the characteristic $\eta=(\eta^1,\dots,\eta^q)$ of the vector field $Q$ written in evolutionary form, which is given by $\eta^\alpha=\phi^\alpha-\xi^i u^\alpha_i$ is also a set of $q$ local CL multipliers of the associated Euler--Lagrange equations of $\mathcal S[u]$, i.e.\ $\eta=\Lambda$. This means that a set of local conservation CL multipliers satisfies the determining equations for local symmetries, in evolutionary form, of the corresponding Euler--Lagrange equations.

In practice, the Euler--Lagrange equations follow from taking the variational derivative of~\eqref{eq:ActionFunctional}, which boils down to applying the Euler operator $\mathsf E_u$ to the Lagrangian function~$L$. The following definitions and theorem will be useful for the construction of \emph{variational parameterization schemes}, i.e.\ parameterization schemes that preserve the variational structure of a system of partial differential equations. A more comprehensive discussion on the relation between symmetries, CLs and variational forms can be found in~\cite{blum89Ay,blum10Ay,olve86Ay}.

\begin{definition}
 Let there be given a system of differential equations $\mathcal L\colon \Delta_l(x,u^{(n)})=0$, $l=1,\dots,L$. The \emph{linearizing operator} (\emph{Fr\'{e}chet derivative}) associated with $\mathcal L$ is the matrix-valued differential operator $\mathsf{D}_{\mathcal L}$ whose components are given by
 \[
  (\mathsf{D}_{\mathcal L})_{\mu\nu}=\left(\pdl{\Delta_\mu}{u^\nu} + \pdl{\Delta_\mu}{u^\nu_{j_1}}\DD_{j_1}+\cdots+\pdl{\Delta_\mu}{u^\nu_{j_1\cdots j_p}}\DD_{j_1}\cdots\DD_{j_p}\right)=\pdl{\Delta_\mu}{u^\nu_J}\DD_J.
 \]
\end{definition}

\begin{definition}
 The \emph{adjoint} $\mathsf{D}^*_{\mathcal L}$ of the linearizing operator $\mathsf{D}_{\mathcal L}$ is the matrix-valued differential operator whose components are given by
 \[
  (\mathsf{D}^*_{\mathcal L})_{\nu\mu}V^\mu=(-\DD)_J\left(\pdl{\Delta_\mu}{u^\nu_J}V^\mu\right).
 \]
 for any differential function $V=(V^1,\dots,V^L)$, where $V^l=V^l(x,u^{(n)})$. The operator $\mathsf{D}_{\mathcal L}$ is \emph{self-adjoint} if and only if $\mathsf{D}^*_{\mathcal L}=\mathsf{D}_{\mathcal L}$.
\end{definition}

\begin{theorem}\label{thm:OnVariationalSystems}
 Let $\mathcal L\colon \Delta_l(x,u^{(n)})=0$, $l=1,\dots,L$ be a system of differential equations. The system $\mathcal L$ can be derived from a variational principle~\eqref{eq:ActionFunctional} if and only if the linearization operator $\mathsf{D}_{\mathcal L}$ associated with $\mathcal L$ is self-adjoint, i.e.\ $\mathsf{D}^*_{\mathcal L}=\mathsf{D}_{\mathcal L}$.
\end{theorem}

Theorem~\ref{thm:OnVariationalSystems} is the key for the construction of variational parameterization schemes. Suppose that the system $\mathcal L$ admits a variational form~\eqref{eq:ActionFunctional}. Averaging of the equations from system $\mathcal L$ and assuming a general parameterization ansatz~\eqref{eq:GeneralFormOfParameterizationScheme} leads to a system of equations that can be brought into the form
\begin{equation}\label{eq:SolvedFormOfParameterizationScheme}
 \Delta_l (x,\bar u^{(n)})=g_l(x,\bar u^{(r)}),\quad l=1,\dots,L.
\end{equation}
See also Remark~\ref{rem:OnExistenceOfSolutionConservativeSchemesInverse}. Since the linearization operator associated with the left-hand side of the above expression is by supposition self-adjoint, the above closed system of differential equations~\eqref{eq:SolvedFormOfParameterizationScheme} will remain variational if and only if
\begin{equation}\label{eq:ConditionForVariationalParameterizationSchemes}
 \mathsf{D}^*_g=\mathsf{D}_g,
\end{equation}
i.e.\ the linearization operator associated with the right-hand sides of system~\eqref{eq:SolvedFormOfParameterizationScheme} must be self-adjoint. This imposes the required conditions on the function $g=(g^1,\dots,g^L)$ to yield a variational parameterization scheme.

Since the one-dimensional shallow-water equations~\eqref{eq:ShallowWater}, which will be our running example for finding conservative parameterization schemes, are not derivable from a variational principle, we illustrate the idea of variational parameterization schemes for the potential Korteweg--de Vries equation.

\begin{example}
 The potential Korteweg--de Vries (KdV) equation is obtained from the usual KdV equation $v_t+vv_x+v_{xxx}=0$ by the differential substitution $v=u_x$, i.e.
 \begin{equation}\label{eq:PotentialKdV}
  \Delta_{\rm pKdV}(x,u^{(4)})=u_{tx}+u_xu_{xx}+u_{xxxx}=0.
 \end{equation}
 This system is variational with $L=\frac12u_{xx}^2-\frac16u_x^3-\frac12u_tu_x$ being the Lagrangian. The maximal set of point symmetries of~\eqref{eq:PotentialKdV} is infinite dimensional and spanned by the vector fields
 \[
  \p_t,\quad \p_x,\quad t\p_x+x\p_u,\quad \gamma(t)\p_u,\quad 3t\p_t+x\p_x-u\p_u,
 \]
 where $\gamma=\gamma(t)$ runs through the set of smooth functions of $t$. The first four vector fields satisfy the variational symmetry condition~\eqref{eq:ConditionOnVariationalSymmetries} and are thus associated with local parameterization laws. The characteristics of these symmetries are (i) $\eta=-u_t$, (ii) $\eta=-u_x$, (iii) $\eta=x-tu_x$ and (iv) $\eta=\gamma(t)$ to which the following CLs $\Lambda(x,U,U_x,U_{xx},U_{xxx})\Delta_{\rm pKdV}(x,U^{(4)})=\DD_t\rho+\DD_xF$ correspond~\cite{blum10Ay}:
\begin{align*}
\begin{array}{lll}
 \hline
 \multicolumn{1}{c}{\Lambda} & \multicolumn{1}{c}{\rho} & \multicolumn{1}{c}{F}\\
 \hline\noalign{\smallskip}
 U_x^2+2U_{xxx} &\quad \frac13U_x^3+U_xU_{xxx} &\quad \frac14U_x^4+ U_x^2U_{xxx}-U_xU_{txx} + U_{xxx}^2+U_{xx}U_{tx},\\
 U_x &\quad \frac{1}{2}U_x^2 &\quad \frac13U_x^3+U_xU_{xxx}-\frac12U_{xx}^2,\\
 x-tU_x &\quad \frac12tU_x^2+xU_x &\quad -\frac13tU_x^3-tU_xU_{xxx}+\frac12tU_{xx}^2+\frac12xU_x^2-U_{xx}+xU_{xxx}, \\
 \gamma(t) &\quad \gamma U_x &\quad \frac12\gamma U_x^2+\gamma U_{xxx}-\gamma_tU.\\
 \hline
\end{array}
\end{align*}

We Reynolds average the potential KdV equation to obtain $\bar u_{tx}+\bar u_x\bar u_{xx}+\bar u_{xxxx}=-\overline{u_x'u_{xx}'}$. To keep the computations as simple as possible we aim at finding the unclosed term $\overline{u_x'u_{xx}'}$ as a function of $u_x$ and $u_{xx}$ in such a manner that the closed equations from the class
 \[
  u_{tx}+u_xu_{xx}+u_{xxxx}=g(u,u_x,u_{xx}),
 \]
are still derivable from a variational principle. From now on we omit bars over $u$ to simplify the notation. The linearization operator associated with the right-hand side of this closed class of potential KdV equations is given by $\mathsf{D}=g_u+g_{u_x}\DD_x+g_{u_{xx}}\DD_x^2$, which has the adjoint $\mathsf{D}^*=g_{u_{xx}}\DD_x^2+(2\DD_xg_{u_{xx}}-g_{u_x})\DD_x + \DD_x^2 g_{u_{xx}}-\DD_x g_{u_x}+g_u$. Thus, condition~\eqref{eq:ConditionForVariationalParameterizationSchemes} yields the system
 \[
  \DD_xg_{u_{xx}}-g_{u_x}=0,\quad \DD_x^2g_{u_{xx}}-\DD_xg_{u_x}=0
 \]
that must be satisfied by function $g$ for an equation from the closed class of potential KdV equations to remain variational. Solving this system of differential equations yields $g=(c_2u_x+c_1)u_{xx}+g^1(u)$, where $c_1$ and $c_2$ are arbitrary constants and $g^1$ is an arbitrary function of $u$. Hence
  \begin{equation}\label{eq:PotentialKdVClosed}
  u_{tx}+u_xu_{xx}+u_{xxxx}-(c_2u_x+c_1)u_{xx}-g^1=0,
 \end{equation}
 is the only admissible form of equations from the class of closed potential KdV equations that is an Euler--Lagrange equation. The associated Lagrangian of~\eqref{eq:PotentialKdVClosed} is given by
 \[
  L=\frac12u_{xx}^2-\frac16(1-c_2)u_x^3-\frac12u_tu_x-\frac{c_1}{2}u_x^2-G(u),
 \]
where $G(u)=\int g^1\ddd u$. For arbitrary $g^1$, the only point symmetries admitted by Eq.~\eqref{eq:PotentialKdVClosed} are generated by the vector fields $\p_t$ and $\p_x$, which are variational and give rise to two local CLs of~\eqref{eq:PotentialKdVClosed}.
\end{example}

\section{Conservative closure schemes for the shallow-water equations}\label{sec:ConservativeParameterizationsShallowWater}

In this section we construct conservative parameterization schemes for the one-dimensional system of shallow-water equations. In nondimensional form, this system reads as
\begin{align}\label{eq:ShallowWater}
\begin{split}
 &\Delta_1(t,x,u^{(1)},h^{(1)})=u_t+uu_x+h_x=0,\\
 &\Delta_2(t,x,u^{(1)},h^{(1)})=h_t+uh_x+hu_x=0,
\end{split}
\end{align}
where $u$ is the velocity and $h$ is the height of the water column over a fixed reference level. For the sake of simplicity, we assume a flat bottom topography in which case $h$ can be considered as the total height of the water column. The first equation is the momentum equation, the second equation the shallow-water continuity equation.

Reynolds-averaging the above system~\eqref{eq:ShallowWater} with the averaging interval $T$ being finite and following the averaging rule for products, $\overline{u^iu^j} = \bar u^i\bar u^j + \overline{u^{i'}u^{j'}}$, then leads to the averaged shallow-water equations
\begin{align}\label{eq:ShallowWaterAveraged}
\begin{split}
 &\bar u_t+\bar u\bar u_x+\bar h_x=-\frac12(\overline{u'^2})_x,\\
 &\bar h_t+\bar u\bar h_x+\bar h\bar u_x=-(\overline{h'u'})_x.
\end{split}
\end{align}
The right-hand sides of the above system are the subgrid-scale quantities that must be parameterized, i.e.\ it is necessary to find a functional relation that allows one to express these terms using only the grid-scale quantities, i.e.\
\begin{equation*}
 -\frac12(\overline{u'^2})_x=f(t,x,\bar u^{(r_1)}, \bar h^{(q_1)}),\quad -(\overline{h'u'})_x=g(t,x,\bar u^{(r_2)}, \bar h^{(q_2)}),
\end{equation*}
where $r_1, r_2, q_1, q_2\in \mathbb{N}_0$. It is the purpose of this section to give two examples of forms of the parameterization functions $f$ and $g$ that lead to closed equations
\begin{align}\label{eq:ShallowWaterClosedGeneralForm}
 u_t+uu_x+h_x=f,\quad h_t+uh_x+hu_x=g
\end{align}
possessing certain conservation laws. Here and in what follows, we omit the bars over averaged variables in the closed class of shallow-water equations since there is no risk of confusion as only averaged forms of dependent variables and their derivatives arise in such equations.

This notation is also convenient since in several of the examples given below, we consider the related problem of finding right-hand sides $f$ and $g$ in system~\eqref{eq:ShallowWaterClosedGeneralForm} preserving certain CLs of system~\eqref{eq:ShallowWater}, i.e.\ we consider structure-preserving extensions of the shallow-water equations as discussed in Remark~\ref{rem:OnExtensionsOfDifferentialEquations}.

Before determining ansatzes for $f$ and $g$ that possess CLs or preserve certain CLs holding for the free shallow-water equations~\eqref{eq:ShallowWaterAveraged} it is instructive to determine the CLs of system~\eqref{eq:ShallowWater}. We restrict ourselves to multipliers that depend on $t$, $x$, $U$ and $H$, i.e.~$\Lambda^1=\Lambda^1(t,x,U,H)$ and $\Lambda^2=\Lambda^2(t,x,U,H)$, where $U$ and $H$ are arbitrary functions of $t$ and $x$. The determining equations~\eqref{eq:DeterminingEquationsMultipliers} for multipliers $\Lambda^1$ and $\Lambda^2$ in this case become
\begin{align}\label{eq:DeterminingEquationsMultipliersShallowWater}
\begin{split}
 &\mathsf E_U(\Lambda^1\Delta_1(t,x,U^{(1)},H^{(1)})+\Lambda^2\Delta_2(t,x,U^{(1)},H^{(1)}))\equiv0,\\
 &\mathsf E_H(\Lambda^1\Delta_1(t,x,U^{(1)},H^{(1)})+\Lambda^2\Delta_2(t,x,U^{(1)},H^{(1)}))\equiv0,
\end{split}
\end{align}
where the Euler operators $\mathsf E_U$ and $\mathsf E_H$ are given by
\[
 \mathsf E_U=\p_U-\DD_t\p_{U_t}-\DD_x\p_{U_x}+\cdots,\quad \mathsf E_H=\p_H-\DD_t\p_{H_t}-\DD_x\p_{H_x}+\cdots.
\]
Splitting the system~\eqref{eq:DeterminingEquationsMultipliersShallowWater} with respect to the derivatives of $U$ and $H$, one obtains the following system of determining equations for CL multipliers
\begin{align}\label{eq:DeterminingEquationsMultipliersShallowWaterExpanded}
\begin{split}
 &\Lambda^1_H-\Lambda^2_U=0,\quad \Lambda^1_U-H\Lambda^2_H=0,\quad \Lambda^2_t+U\Lambda^2_x+\Lambda^1_x=0,\quad \Lambda^1_t+U\Lambda^1_x+H\Lambda^2_x=0.
\end{split}
\end{align}
Differentiating the third equation in system~\eqref{eq:DeterminingEquationsMultipliersShallowWaterExpanded} with respect to~$H$ and multiplying the resulting equation with $H$, one obtains, upon recombining with the first two equations, the equation
\begin{equation}\label{eq:DeterminingEquationsMultipliersShallowWaterExpandedDerived}
\Lambda^1_{Ut}+U\Lambda^1_{Ux}+H\Lambda^2_{Ux}=0.
\end{equation}
Differentiating the last equation in system~\eqref{eq:DeterminingEquationsMultipliersShallowWaterExpanded} with respect to $U$ and then combining the resulting equation with~\eqref{eq:DeterminingEquationsMultipliersShallowWaterExpandedDerived}, one finds that $\Lambda^1_x=0$. Then differentiating the second equation in~\eqref{eq:DeterminingEquationsMultipliersShallowWaterExpanded} with respect to $x$ one gets $\Lambda^2_{Hx}=0$ and thus $\Lambda^2_{Ht}=0$ in view of the third equation in system~\eqref{eq:DeterminingEquationsMultipliersShallowWaterExpanded}. Differentiation of the fourth equation in~\eqref{eq:DeterminingEquationsMultipliersShallowWaterExpanded} with respect to $x$ yields $\Lambda^2_{xx}=0$ and hence $\Lambda^2_{tx}=0$ due to the third equation.

Using these results, the integration of system~\eqref{eq:DeterminingEquationsMultipliersShallowWaterExpanded} leads to
\begin{align}\label{eq:MultipliersShallowWater}
\begin{split}
 &\Lambda^1=-c_1tH+\lambda^1(U,H),\\
 &\Lambda^2=c_1(x-tU)+\lambda^2(U,H),
\end{split}
\end{align}
where $c_1=\const$ and $\lambda^1=\lambda^1(U,H)$ and $\lambda^2=\lambda^2(U,H)$ are any functions satisfying the system $\lambda^1_H-\lambda^2_U=0$ and $\lambda^1_U-H\lambda^2_H=0$. Hence there are an infinite number of associated CLs, which reflects the possibility of linearizing the quasilinear system~\eqref{eq:ShallowWater} using a hodograph transformation (interchanging the dependent and independent variables), see also Section~\ref{sec:ShallowWaterInvariantConservative}. More details on the connection between CLs and linearization of partial differential equations can be found in~\cite{anco08Ay,blum10Ay}.

\subsection{Conservative parameterization schemes via direct classification}\label{sec:ShallowWaterDirectMethod}

In this subsection we give an example for the construction of conservative parameterization schemes using the technique of direct CL classification.

As an example, consider the problem of finding diffusion terms of the form
\begin{align}\label{eq:ShallowWaterDissipationDirectMethod}
\begin{split}
 &u_t + uu_x + h_x - F(h,u_x,h_x)u_{xx}=0,\\
 &h_t + uh_x + hu_x=0,
\end{split}
\end{align}
allowing for CLs arising from multipliers of the form $\Lambda^1=\Lambda^1(t,x,U,H)$ and $\Lambda^2=\Lambda^2(t,x,U,H)$, where $U$ and $H$ are arbitrary functions of $t$ and $x$. The problem is to first determine corresponding CLs arising for arbitrary $F$. Following this, we determine particular forms of $F$ that yield additional CLs. The classification is done up to equivalence.

\begin{theorem}
The equivalence algebra $\mathfrak g^\sim$ of the class of one-dimensional dissipative shallow-water equations is generated by the following basis elements
\begin{equation}\label{eq:EquivalenceAlgebraDissipativeShallowWater}
 \p_t,\quad \p_x,\quad t\p_x+\p_u,\quad t\p_t-u\p_u-2h\p_h-F\p_F,\quad x\p_x+u\p_u+2h\p_h+2F\p_F.
\end{equation}
Besides the associated continuous equivalence transformations~\eqref{eq:EquivalenceAlgebraDissipativeShallowWater}, the class of equations~\eqref{eq:ShallowWaterDissipationDirectMethod} admits two independent discrete equivalence transformations, which are given by $(t,x,u,h,F)\mapsto(-t,-x,u,h,-F)$ and $(t,x,u,h,F)\mapsto(-t,x,-u,h,-F)$, respectively. The continuous and discrete equivalence transformations form the equivalence group $G^\sim$ of the class~\eqref{eq:ShallowWaterDissipationDirectMethod}.
\end{theorem}

\noindent The determining equations for CL multipliers $\Lambda^1$ and $\Lambda^2$ are given by
\begin{align}\label{eq:ShallowWaterDeterminingEquationsDirectMethod}
\begin{split}
 &\Lambda^1_H-\Lambda^2_U=0,\quad \Lambda^1_U - H\Lambda^2_H=0,\quad \Lambda^1_x+U\Lambda^2_x+\Lambda^2_t=0,\\
 &\Lambda^1 F_{H_x}=0,\quad H_x\Lambda^1F_{HU_x}+(U_x\Lambda^1_U+H_x\Lambda^1_H+\Lambda^1_x)F_{U_x}+2\Lambda^1_UF=0,\\
 &2H_x\Lambda^1F_{HH_x}+2(U_x\Lambda^1_U+H_x\Lambda^1_H+\Lambda^1_x)F_{H_x}+\Lambda^1F_H+F\Lambda^1_H=0,\\
 &H_x\Lambda^1F_{HH_x}+(\Lambda^1_x+U_x\Lambda^1_U+H_x\Lambda^1_H)F_{H_x}-\Lambda^1F_H-\Lambda^1_HF=0,\\
 &H_x^2\Lambda^1F_{HH}+2(U_xH_x\Lambda^1_U+H_x^2\Lambda^1_H+H_x\Lambda^1_x)F_H+(U_x^2\Lambda^1_{UU}+2U_xH_x\Lambda^1_{UH}+{}\\
 &{}H_x^2\Lambda^1_{HH}+2U_x\Lambda^1_{Ux}+2H_x\Lambda^1_{Hx}+\Lambda^1_{xx})F+U\Lambda^1_x+H\Lambda^2_x+\Lambda^1_t=0.
\end{split}
\end{align}

Assuming $F$ to be arbitrary, one can split the above system with respect to the various derivatives of $F$, which then leads to the solution $\Lambda^1=0$ and $\Lambda^2=\const$. \emph{That is, for arbitrary $F$, the only CL admitted by system~\eqref{eq:ShallowWaterDissipationDirectMethod}, corresponding to the specified class of multipliers, is conservation of mass.}

We now consider particular forms of $F$ for which system~\eqref{eq:ShallowWaterDissipationDirectMethod} possesses additional CLs. From the classifying condition $\Lambda^1F_{H_x}=0$ in system~\eqref{eq:ShallowWaterDeterminingEquationsDirectMethod} it follows that $\Lambda^1=0$ when $F_{H_x}\ne0$. In this case, no CL extension exists. Thus, we only study the case of $F_{H_x}=0$ subsequently. In this case, the system of determining equations~\eqref{eq:ShallowWaterDeterminingEquationsDirectMethod} simplifies significantly since all terms involving derivatives of $F_{H_x}$ vanish. Consequently, it is possible to split the resulting equations with respect to the powers of $H_x$. The resulting system of determining equations is given by
\begin{subequations}
\begin{align}
 &\Lambda^1_H-\Lambda^2_U=0,\quad \Lambda^1_U - H\Lambda^2_H=0,\quad \Lambda^1_x+U\Lambda^2_x+\Lambda^2_t=0,\label{eq:ShallowWaterDeterminingEquationsDirectMethodGeneral} \\
 &\Lambda^1F_H+F\Lambda^1_H=0,\label{eq:ShallowWaterDeterminingEquationsDirectMethodDet1}\\
 &(U_x\Lambda^1_U+\Lambda^1_x)F_{U_x}+2\Lambda^1_UF=0,\label{eq:ShallowWaterDeterminingEquationsDirectMethodDet2}\\
 &(U_x^2\Lambda^1_{UU}+2U_x\Lambda^1_{Ux}+\Lambda^1_{xx})F+U\Lambda^1_x+H\Lambda^2_x+\Lambda^1_t=0.\label{eq:ShallowWaterDeterminingEquationsDirectMethodDet3}
\end{align}
\end{subequations}
Equations~\eqref{eq:ShallowWaterDeterminingEquationsDirectMethodGeneral} do not involve the constitutive function $F$ and thus can be integrated immediately. This results in
\begin{align}\label{eq:GeneralFormOfMultipliersDirectMethod}
\begin{split}
 &\Lambda^1=-\frac12c_2tU^2+(c_1H+c_2x+c_3)U-c_2tH\ln H-\alpha_1(t)H-\frac12\alpha_1''(t)x^2{}\\
 &-\alpha_2'(t)x+\alpha_3(t),\\
 &\Lambda^2=\frac12c_1U^2-\left(\alpha_1(t)+c_2t\right)U- (c_2(Ut-x)-c_3)\ln H+c_1H  + \alpha_1'(t)x+\alpha_2(t),
\end{split}
\end{align}
where $\alpha_1(t)$, $\alpha_2(t)$ and $\alpha_3(t)$ are arbitrary smooth functions of $t$, a prime denotes the derivative with respect to $t$ and $c_i$, $i=1,\dots,3$, are arbitrary constants.

Each of equations~\eqref{eq:ShallowWaterDeterminingEquationsDirectMethodDet1}--\eqref{eq:ShallowWaterDeterminingEquationsDirectMethodDet3} explicitly involves the constitutive function $F$ and these four equations are solved using compatibility analysis. Thus, different cases arise.

\medskip

\noindent \textit{Case (I), $F_H=0$.} From Eq.~\eqref{eq:ShallowWaterDeterminingEquationsDirectMethodDet1} it follows that $\Lambda^1_H=0$ and Eq.~\eqref{eq:ShallowWaterDeterminingEquationsDirectMethodGeneral} implies that $\Lambda^2_U=0$. From the form of the multipliers~\eqref{eq:GeneralFormOfMultipliersDirectMethod} we find that $c_1=c_2=0$, $\alpha_1(t)=0$ and thus the multipliers in this case are of the form
\[
 \Lambda^1=c_3U-\alpha_2'(t)x+\alpha_3(t),\quad \Lambda^2=c_3\ln H + \alpha_2(t),
\]
Substituting this into Eq.~\eqref{eq:ShallowWaterDeterminingEquationsDirectMethodDet3} results in
\[
 \alpha_2'(t)U+\alpha_2''(t)x-\alpha_3'(t)=0,
\]
which leads to $\alpha_2(t)=\const=\alpha_2$ and $\alpha_3(t)=\const=\alpha_3$. Eq.~\eqref{eq:ShallowWaterDeterminingEquationsDirectMethodDet2} then reduces to
\[
 c_3(U_xF_{U_x}+2F)=0.
\]
This equation implies that either (i) $c_3=0$ and $F=F(U_x)$ is arbitrary or (ii) $c_3\ne0$ and $F=c_0/U_x^2$, with being the integration constant. Since $c_0\ne0$ by assumption, we can use the transformations from the equivalence group $G^\sim$ to scale $c_0=1$. Thus in subcase (i) there is one additional CL which is of the form
\begin{equation}\label{eq:FirstConservationLawExtension}
 \DD_tu + \DD_x\left(\frac12u^2+h-\int F(u_x)\ddd u_x\right)=0.
\end{equation}
In subcase (ii), the multipliers are
\[
 \Lambda^1=c_3U+\alpha_3,\quad \Lambda^2=c_3\ln H+\alpha_2
\]
so that here there are two additional CLs. The CL associated with $\alpha_3=\const$ is CL~\eqref{eq:FirstConservationLawExtension} provided that $F=1/u_x^2$. The second CL associated with $c_3$ is of the form
\[
 \DD_t\left(\frac12u^2+(h\ln h-h)-t\right)+\DD_x\left(\frac13u^2+h\ln h+\frac{1}{u_x}\right)=0.
\]

\medskip

\noindent \textit{Case (II), $f_H\ne0$.} Substituting $\Lambda^1$ in the form given in Eq.~\eqref{eq:GeneralFormOfMultipliersDirectMethod} into Eq.~\eqref{eq:ShallowWaterDeterminingEquationsDirectMethodDet1} we can split the resulting equation
\begin{align*}
 &\bigg(-\frac12c_2tU^2+(c_1H+c_2x+c_3)U-c_2tH\ln H-\alpha_1(t)H-\frac12\alpha_1''(t)x^2-\alpha_2'(t)x+\alpha_3(t)\bigg)F_H{}\\
 &+\left(c_1U-c_2t(1+\ln H)-\alpha_1(t)\right)F=0,
\end{align*}
with respect to powers of $U$ and $x$ since $F=F(H,U_x)$ only. Splitting with respect to $U^2$, $x^2$ and $x$  implies that $c_2=0$, $\alpha_1(t)=\alpha_1^1t+\alpha_1^0$ and $\alpha_2(t)=\const=\alpha_2$, respectively, where $\alpha_1^1,\alpha_1^0=\const$. Differentiating the simplified equation twice with respect to $t$ leads to $\alpha_3''(t)=0$ or $\alpha_3(t)=\alpha_3^1t+\alpha_3^0$, $\alpha_3^1,\alpha_3^0=\const$. The above equation thus simplifies~to
\begin{align}\label{eq:ShallowWaterDeterminingEquationsDirectMethodDet21}
 &\left((c_1H+c_3)U-(\alpha_1^1t+\alpha_1^0)H+\alpha_3^1t+\alpha_3^0\right)F_H+\left(c_1U
 -\alpha_1^1t-\alpha_1^0\right)F=0.
\end{align}
Plugging the simplified form of $\Lambda^1$ and $\Lambda^2$ into Eq.~\eqref{eq:ShallowWaterDeterminingEquationsDirectMethodDet3} leads to $\alpha_3^1=0$. Splitting Eq.~\eqref{eq:ShallowWaterDeterminingEquationsDirectMethodDet21} with respect to $t$ and $U$ yields the following system of three equations
\begin{subequations}\label{eq:ShallowWaterDeterminingEquationsDirectMethodDetMaxSimplification}
\begin{equation}\label{eq:ShallowWaterDeterminingEquationsDirectMethodDetMaxSimplification1}
 \alpha_1^1HF_H+\alpha_1^1F=0,\quad (c_1H+c_3)F_H+c_1F=0,\quad (\alpha_1^0H-\alpha_3^0)F_H+\alpha_1^0F=0.
\end{equation}
At the same time, the remaining classifying equation~\eqref{eq:ShallowWaterDeterminingEquationsDirectMethodDet2} gives
\begin{equation}
 (c_1H+c_3)(2F+U_xF_{U_x})=0.
\end{equation}
\end{subequations}
System~\eqref{eq:ShallowWaterDeterminingEquationsDirectMethodDetMaxSimplification} allows one to find four inequivalent solutions. One either has (i) $(c_1,c_3,\alpha_1^1)=(0,0,0)$, (ii) $(c_1,c_3)\ne(0,0)$, $\alpha_1^1=0$ and $\alpha_1^0=\delta c_1$, $\alpha_3^0=-\delta c_3$, (iii) $(c_1,c_3,\alpha_3^0)=(0,0,0)$ and $\alpha_1^0=-\delta \alpha_1^1$ or (iv) $(c_3,\alpha_3^0)=(0,0)$, $c_1\ne0$ and $\alpha_1^1=\delta c_1$, $\alpha_1^0=\varepsilon c_1$, $\delta,\epsilon=\const$. In all other cases $F=0$, which is excluded from consideration. In subcase (i), the classifiyng equations~\eqref{eq:ShallowWaterDeterminingEquationsDirectMethodDetMaxSimplification} are integrated to give
\[
 F=\frac{g(U_x)}{\alpha_1^0H-\alpha_3^0},
\]
where $g=g(U_x)$ is an arbitrary non-vanishing smooth function of $U_x$ and $\alpha_1^0\ne0$ since otherwise the assumption $F_H\ne0$ would be contradicted. In this subcase, the multipliers $\Lambda^1$ and $\Lambda^2$ are given by
\[
 \Lambda^1=\alpha_1^0H-\alpha_3^0,\quad \Lambda^2=\alpha_1^0U+\alpha_2.
\]
The two CLs that correspond to these characteristics are conservation of mass, which follows from the multipliers $(\Lambda^1,\Lambda^2)=(0,\alpha_2)$ and
\[
 \DD_t(\alpha_1^0hu-\alpha_3^0u)+\DD_x\left(\frac12\alpha_1^0 h^2-\frac12\alpha_3^0u^2+\alpha_1^0hu^2-\alpha_3^0h-\int g(u_x)\ddd u_x\right)=0,
\]
for the multipliers $(\Lambda^1,\Lambda^2)=(\alpha_1^0H-\alpha_3^0,\alpha_1^0U)$.

In subcase (ii), the solution of~\eqref{eq:ShallowWaterDeterminingEquationsDirectMethodDetMaxSimplification} is $F=c_0U_x^{-2}/(c_1H+c_3)$ and using the equivalence transformations from $G^\sim$ one can put $c_0=1$. The multipliers $\Lambda^1$ and $\Lambda^2$ are of the form
\[
 \Lambda^1=(c_1H+c_3)(U-\delta),\quad \Lambda^2=\frac12c_1U^2-\delta c_1U+c_3\ln H+c_1H+\alpha_2.
\]
The CLs associated with the found multipliers are again conservation of mass associated with $(\Lambda^1,\Lambda^2)=(0,\alpha_2)$ and
\[
 \DD_t(c_1hu+c_3u)+\DD_x\left(\frac12(c_3u^2+c_1h^2)+(c_1u^2+c_3)h+\frac{1}{u_x}\right)=0
\]
for the multiplier $(\Lambda^1,\Lambda^2)=(c_1H+c_3,c_1U)$. The third CL, corresponding to the multipliers $(\Lambda^1,\Lambda^2)=((c_1H+c_3)U,c_1U^2/2+c_3\ln H+c_1H)$, is
\begin{align*}
 &\DD_t\left(\frac12(c_1h+c_3)u^2+c_3(h\ln h-h)+\frac12c_1h^2-t\right)+\DD_x\bigg(\frac12c_1hu^3+\frac13c_3u^3+{}\\
 &c_3hu\ln h + c_1h^2u+\frac{u}{u_x}\bigg)=0.
\end{align*}

In subcase (iii) the solution of system~\eqref{eq:ShallowWaterDeterminingEquationsDirectMethodDetMaxSimplification} is $F=H^{-1}g(U_x)$, for an arbitrary non-vanishing smooth function $g(U_x)$. Here the multipliers are of the form
\[
 \Lambda^1=\alpha_1^1(\delta-t)H,\quad \Lambda^2=\alpha_1^1(x-Ut)+\delta a_1^1U+\alpha_2.
\]
The three CLs associated with these multipliers are again conservation of mass ($\Lambda^1=0,\Lambda^2=\alpha_2$), the CL
\[
 \DD_t(xh-thu)+\DD_x\left(xhu-thu^2-\frac12th^2+t\int g(u_x)\ddd u_x\right)=0,
\]
corresponding to $(\Lambda^1,\Lambda^2)=(-tH,x-Ut)$ and the CL
\[
 \DD_t(hu)+\DD_x\left(\frac12h^2+hu^2-\int g(u_x)\ddd u_x\right)=0
\]
which stems from the multipliers $(\Lambda^1,\Lambda^2)=(H,U)$.

In the final subcase (iv) we obtain $F=c_0H^{-1}U_x^{-2}$ from the integration of the system~\eqref{eq:ShallowWaterDeterminingEquationsDirectMethodDetMaxSimplification}. Again $c_0=1$ mod $G^\sim$. The multipliers in this case are
\[
 \Lambda^1= (c_1(U-\delta t)-\epsilon c_1)H,\quad \Lambda^2=\frac12c_1U^2+c_1H+\delta c_1(x-Ut)-\epsilon c_1U+\alpha_2.
\]
Besides the obvious conservation of mass, the three other CLs include
\[
 \DD_t(xh-thu)+\DD_x\left(xhu-thu^2-\frac12th^2-\frac{t}{u_x}\right)=0,
\]
stemming from $(\Lambda^1,\Lambda^2)=(-tH,x-Ut)$,
\[
 \DD_t(hu)+\DD_x\left(\frac12h^2+hu^2+\frac{1}{u_x}\right)=0
\]
which is obtained from the multipliers $(\Lambda^1,\Lambda^2)=(H,U)$, and conservation of energy
\[
 \DD_t\left(\frac12hu^2+\frac12h^2-t\right)+\DD_x\left(\frac{1}{2}hu^3+h^2u+\frac{u}{u_x}\right)=0,
\]
associated with the multipliers $(\Lambda^1,\Lambda^2)=(UH,U^2/2+H)$

\medskip

\noindent Comparing the results of the two cases $F_H=0$ and $F_H\ne0$ we have proved the following.

\begin{theorem}
 For $F\ne0$ any equation from the class of dissipative systems of one-dimensional shallow-water equations~\eqref{eq:ShallowWaterDissipationDirectMethod} has at most four linearly independent conservation laws arising from multipliers of the form $\Lambda^1=\Lambda^1(t,x,U,H)$ and $\Lambda^2=\Lambda^2(t,x,U,H)$. A complete list of $G^\sim$-inequivalent equations and their associated conservation laws is given in Table~\ref{tab:ConservationLawsDirectMethod},
 where $g=g(u_x)$ is an arbitrary non-vanishing smooth function of $u_x$ and $c_1$ and $c_2$ are arbitrary constants with $c_2\ne0$.
 \begin{table}[ht!]
 \centering
 \caption{Conservation law classification of a class of one-dimensional dissipative shallow-water equations}
 \renewcommand\arraystretch{1.5}
 \begin{tabular}{ll}
 \hline
 Form of $F$ & Conservation laws\\
 \hline
 \hline
 $\forall F$ & $\textup{CL}_1=\DD_t h + \DD_x(hu)=0$\\
 \hline
 \multirow{2}{*}{$F=\dfrac{g(u_x)}{(c_1h+c_2)}$} & $\textup{CL}_1=0$,\\ & $\DD_t(c_1hu+c_2u)+\DD_x\left(\dfrac12c_1h^2+\dfrac12c_2u^2+c_1hu^2+c_2-\int g(u_x)\ddd u_x\right)=0$\\
 \hline
 \multirow{4}{*}{$F=\dfrac{1}{(c_1h+c_2)u_x^2}$} & $\textup{CL}_1=0$,\\ & $\DD_t(c_1hu+c_2u)+\DD_x\left(\dfrac12(c_2u^2+c_1h^2)+(c_1u^2+c_2)h+\dfrac{1}{u_x}\right)=0$,\\ & $\DD_t\left(\dfrac12(c_1h+c_2)u^2+c_2(h\ln  h-h)+\dfrac12c_1h^2-t\right)+$\\ & $\DD_x\bigg(\dfrac12c_1hu^3+\dfrac13c_2u^3+c_2hu\ln h + c_1h^2u+\dfrac{u}{u_x}\bigg)=0$\\
 \hline
 \multirow{3}{*}{$F=\dfrac{g(u_x)}{h}$} & $\textup{CL}_1=0$,\\
 & $\DD_t(xh-thu)+\DD_x\left(xhu-thu^2-\frac12th^2+t\int g(u_x)\ddd u_x\right)=0$,\\
 & $\DD_t(hu)+\DD_x\left(\frac12h^2+hu^2-\int g(u_x)\ddd u_x\right)=0$ \\
 \hline
 \multirow{4}{*}{$F=\dfrac{1}{hu_x^2}$} & $\textup{CL}_1=0$,\\
 &  $\DD_t(xh-thu)+\DD_x\left(xhu-thu^2-\dfrac12th^2-\dfrac{t}{u_x}\right)=0,$ \\
 &  $\DD_t(hu)+\DD_x\left(\dfrac12h^2+hu^2+\dfrac{1}{u_x}\right)=0,$\\
 & $\DD_t\left(\dfrac12hu^2+\dfrac12h^2-t\right)+ \DD_x\left(\dfrac12hu^3+h^2u+\dfrac{u}{u_x}\right)=0$\\
 \hline
 \end{tabular}
 \label{tab:ConservationLawsDirectMethod}
 \end{table}

\end{theorem}

\subsection{Conservative parameterization schemes via inverse classification}

In this subsection, two examples are presented that illustrate the procedure for finding conservative parameterization schemes through inverse group classification.

\medskip

\noindent\textbf{Parameterizations conserving energy, mass and momentum.} In this example we focus on four physical CLs, namely conservation of mass-specific momentum, mass, momentum and energy which correspond to the multipliers
\begin{equation}\label{eq:MultipliersShallowWaterSub}
 \Lambda^1=c_1+c_3H+c_4UH,\quad \Lambda^2=c_2+c_3U+c_4\left(\frac12U^2+H\right),
\end{equation}
for $c_1,\dots,c_4\in\mathbb{R}$ for the shallow-water equations~\eqref{eq:ShallowWater}. It can be checked that the above multipliers~\eqref{eq:MultipliersShallowWaterSub} satisfy the system of multiplier determining equations~\eqref{eq:DeterminingEquationsMultipliersShallowWaterExpanded} and thus yield CLs for the shallow-water equations~\eqref{eq:ShallowWater}. The canonical forms $\DD_t\rho+\DD_xX=0$ of the associated CLs are
\begin{align*}
\begin{array}{llll}
 c_1\colon & \quad \rho=u, & X=\frac12u^2+h & \quad \textup{mass-specific momentum},\\
 c_2\colon & \quad \rho=h, & X=uh & \quad \textup{mass}\\
 c_3\colon & \quad \rho=uh, & X=u^2h+\frac12h^2 & \quad \textup{momentum}\\
 c_4\colon & \quad \rho=\frac12(u^2h+h^2) & X=\frac12hu^3+h^2u & \quad \textup{energy}.
\end{array}
\end{align*}

We now consider the problem of finding parameterization schemes~\eqref{eq:ShallowWaterClosedGeneralForm} that preserve the above four multipliers. For the sake of demonstration we limit ourselves to constitutive functions of the form
\[
 f=f(x,u,h,u_x,h_x),\quad g=g(x,u,h,u_x,h_x),
\]
where here and in what follows we omit the bars and the averaging of the dependent variables is to be understood. That is, we look for functions $f$ and $g$ that satisfy
\begin{align}\label{eq:ConservationLawsShallowWaterAveraged}
 \Lambda^1(\Delta_1-f)+\Lambda^2(\Delta_2-g)=\mathrm{D}_t\rho+\mathrm{D}_xX_0,
\end{align}
for the four multipliers~\eqref{eq:MultipliersShallowWaterSub} where $\Delta_1=u_t+uu_x+h_x$, $\Delta_2=h_t+uh_x+hu_x$ as before. Note that adding input terms to the shallow-water equations~\eqref{eq:ShallowWater} will in general lead to modified expressions for $X$, which is why we use $X_0$ in the CL~\eqref{eq:ConservationLawsShallowWaterAveraged}. However, the conserved quantity $\rho$ remains unchanged since the parameterization functions do not depend explicitly on $t$ or derivatives of the unknown functions with respect to $t$.

Applying separately the Euler operators $\mathsf{E}_U$ and $\mathsf{E}_H$ with respect to $U$ and $H$ of Eq.~\eqref{eq:ConservationLawsShallowWaterAveraged}, one obtains the system of determining equations for CL multipliers. Since the multipliers $\Lambda^1$ and $\Lambda^2$ are already prescribed, the resulting system
\[
 \mathsf{E}_U(\Lambda^1(\Delta_1-f)+\Lambda^2(\Delta_2-g))=0,\quad \mathsf{E}_H(\Lambda^1(\Delta_1-f)+\Lambda^2(\Delta_2-g))=0,
\]
is now the system of the determining equations for the parameterization functions $f$ and $g$. The determining equations can be split with respect to the constants $c_1,c_2,c_3,c_4$ and the unconstrained variables which are $t$, $U_t$, $H_t$, $U_{tx}$, $H_{tx}$, $U_{xx}$ and $H_{xx}$. Splitting with respect to the highest derivatives arising yields the elementary equations
\begin{align*}
 &f_{U_xU_x}=f_{H_xU_x}=f_{H_xH_x}=g_{U_xU_x}=g_{H_xU_x}=g_{H_xH_x}=0,
\end{align*}
which can be integrated to give the following constrained form for the functions $f$ and $g$,
\begin{align*}
 &f=f^1(x,U,H)U_x+f^2(x,U,H)H_x+f^3(x,U,H),\\
 &g=g^1(x,U,H)U_x+g^2(x,U,H)H_x+g^3(x,U,H).
\end{align*}
The remaining determining equations can be split to yield the system
\begin{align}\label{eq:DeterminingEquationsMassMomentumEnergyShallowWater}
\begin{split}
 &f^3=g^3=0,\quad f^1_x=f^2_x=0,\quad g^2=f^1,\\
 &g^1-Hf^2=0,\quad f^1_H-f^2_U=0,\quad f^1_U-Hf^2_H-f^2=0.
\end{split}
\end{align}
System~\eqref{eq:DeterminingEquationsMassMomentumEnergyShallowWater} can be integrated to give the most general form of the functions $f$ and $g$ admitting CLs of interest. In particular, one obtains
\begin{align}\label{eq:MassMomentumEnergyPreservingParameterization}
\begin{split}
 &f= f^1U_x + f^2H_x, \\
 &g= f^2HU_x + f^1H_x.
\end{split}
\end{align}
A particular class of solutions is given by
\begin{align}\label{eq:MassMomentumEnergyPreservingParameterizationAne0}
\begin{split}
 &f^1=(\alpha_1\sin{\sqrt{b}U}+\alpha_2\cos\sqrt{b}U)\left(\alpha_3J_0(2\sqrt{bH})+\alpha_4Y_0(2\sqrt{bH})\right)+\alpha_5,\\
 &f^2=\frac{1}{\sqrt{H}}(\alpha_1\cos\sqrt{b}U-\alpha_2\sin{\sqrt{b}U})\left(\alpha_3J_1(2\sqrt{bH})+\alpha_4Y_1(2\sqrt{bH})\right).
\end{split}
\end{align}
In this solution, $\alpha_1,\dots,\alpha_5,b=\const$, $b>0$ and $J_n$ and $Y_n$ are the Bessel functions of the first and second kind, respectively.

The form~\eqref{eq:MassMomentumEnergyPreservingParameterizationAne0} for $f^1$ and $f^2$ does not lead to a particularly physical parameterization ansatz. Physically more relevant forms for $f$ and $g$ can be found upon imposing other restrictions on these functions, which involves finding another interesting set of solutions of~\eqref{eq:DeterminingEquationsMassMomentumEnergyShallowWater}. An example of such a construction is the subclass of parameterization schemes of the form~\eqref{eq:MassMomentumEnergyPreservingParameterization} which in addition to~\eqref{eq:DeterminingEquationsMassMomentumEnergyShallowWater} satisfies the equation $f^1_U=0$. In this case, the functions $f^1$ and $f^2$ in~\eqref{eq:MassMomentumEnergyPreservingParameterization} are given by
\[
 f^1=\beta_1\ln H + \beta_2,\quad f^2=\frac{\beta_1U+\beta_3}{H},
\]
where $\beta_1,\beta_2,\beta_3$ are arbitrary constants.

\medskip

\noindent\textbf{Conservative momentum dissipation schemes.} As discussed in Remark~\ref{rem:OnExtensionsOfDifferentialEquations}, apart from the problem of finding conservative parameterization schemes another question of physical importance is to construct input terms that preserve some of the geometric structure of the initial system.

We illustrate this idea by constructing the most general dissipation term of the form
\[
 u_t +uu_x+h_x=f(x,h,h_x,u_x,u_{xx}),\quad h_t+uh_x+hu_x=0,
\]
for the shallow-water equations that preserves the multipliers $\Lambda^1=c_0$ and $\Lambda^2=c_1$, i.e.\ we set $f=f(x,h,h_x,u_x,u_{xx})$ and $g=0$ in system~\eqref{eq:ShallowWaterClosedGeneralForm}. We do not aim at conserving momentum or energy in this case, because our aim in this example is to construct a dissipation for the shallow-water equations, which by definition should violate energy and momentum conservation.

Using the same procedure as outlined in the previous example, we find that $f$ should be of the form
\begin{align}\label{eq:DissipationShallowWater}
 f=f^1u_{xx}+\left(\int f^1_h\ddd u_x + f^2\right)h_x+\int\left(\int f^1_{xh}\ddd u_x + f^2_x\right)\ddd h + c u_x + f^3,
\end{align}
where $c\in\mathbb{R}$, $f^1=f^1(x,h,u_x)$, $f^2=f^2(x,h)$ and $f^3=f^3(x)$.

From the form of~\eqref{eq:DissipationShallowWater}, one observes that the requirement of conserving both mass and mass-specific momentum leads to a quasi-linear dissipation scheme, i.e.\ $f$ is linear in terms of $u_{xx}$.

\subsection{Conservative and invariant parameterization}\label{sec:ShallowWaterInvariantConservative}

We now turn to the problem of finding parameterization schemes that are both conservative and preserve certain symmetries of the original (unaveraged) system of differential equations. We illustrate this idea with the two examples of the previous section.

The maximal Lie invariance algebra $\mathfrak g$ of the system of one-dimensional shallow-water equations~\eqref{eq:ShallowWater} is infinite dimensional and has the following basis elements
\begin{align}\label{eq:SymmetriesOneDimensionalShallowWater}
\begin{split}
&t\p_t + x\p_x, \quad x\p_x+u\p_u+2h\p_h,\quad  t\p_x + \p_u,\\
&(2x-6tu)\p_t + (6h-3u^2)t\p_x+(u^2+4h)\p_u+4hu\p_h,\quad \tau(h,u)\p_t+ \zeta(h,u)\p_x,
\end{split}
\end{align}
where the functions~$\tau$ and~$\zeta$ run through the set of solutions of the system
\begin{align*}
\begin{split}
&\zeta_h-u\tau_h+\tau_u=0,\quad \zeta_u-u\tau_u+h\tau_h=0.
\end{split}
\end{align*}
The infinite dimensional part of $\mathfrak g$ indicates the existence of a linearization transformation, which is given through the hodograph transformation, i.e.\ $\tau = t$ and $\zeta=x$ are the new dependent variables and $u$ and $h$ are the new independent variables.

The question of which symmetries one aims to preserve when constructing a (conservative) parameterization scheme should be answered using physical arguments. For example, processes that are to be parameterized in the framework of classical mechanics should be represented in such a manner so as to be invariant with respect to the Galilean group. Choices for subgroups to be preserved by a parameterization scheme can be also motivated from compatibility with certain boundary-value problems, see also the related discussion in~\cite{bihl12By}.

As outlined in Section~\ref{sec:ShallowWaterInvariantConservative}, when constructing parameterization schemes that are required to be both invariant and conservative one can follow two ways, namely using direct or inverse group classification. We now use the examples worked out in the previous section to illustrate both ways.

\medskip

\noindent\textbf{Invariant conservative parameterization using direct classification.} The direct group classification method can be illustrated with the first example from the previous section. Essentially, the only freedom one has left with the parameterization~\eqref{eq:MassMomentumEnergyPreservingParameterization}--\eqref{eq:MassMomentumEnergyPreservingParameterizationAne0} is to set to zero some of the constants $\alpha_1,\dots, \alpha_5$ and investigate which symmetries the resulting systems have. This analysis should be done up to equivalence of the class of equations of the form~\eqref{eq:MassMomentumEnergyPreservingParameterization}--\eqref{eq:MassMomentumEnergyPreservingParameterizationAne0}.

The kernel of the maximal Lie invariance algebra from this class is given by the subalgebra of~$\mathfrak g$ consisting of the basis elements
\[
 \mathfrak g^\cap = \langle t\p_t+x\p_x,\p_t,\p_x \rangle.
\]
It turns out that the only extension of this kernel algebra arises when $\alpha_1=\alpha_2=\alpha_3=\alpha_4=0$ and $\alpha_5=c\ne0$. However, in this case the transformation $\tilde u = u-c$ maps the resulting case to the initial shallow-water equations~\eqref{eq:ShallowWater}, i.e.\ the parameterization becomes trivial. Stated in another way, the constant $\alpha_5$ is inessential for the classification problem of the class~\eqref{eq:MassMomentumEnergyPreservingParameterization}--\eqref{eq:MassMomentumEnergyPreservingParameterizationAne0}, which has the two inequivalent solutions $(\alpha_1,\alpha_2,\alpha_3,\alpha_4)\ne(0,0,0,0)$ and $(\alpha_1,\alpha_2,\alpha_3,\alpha_4)=(0,0,0,0)$. Only the first solution leads to a nontrivial parameterization scheme.

A similar analysis could be carried out with the other parameterization schemes of the form~\eqref{eq:MassMomentumEnergyPreservingParameterization} that can be constructed by finding only particular solutions of the system~\eqref{eq:DeterminingEquationsMassMomentumEnergyShallowWater}. As in the previous case, these classes of parameterizations essentially only depend on certain constants. The classification problem then basically reduces to finding those constants that can be set to zero by a proper transformation of the equation variables (inessential constants) and studying the classification problem with respect to the remaining (essential) constants. This is a straightforward task and is not considered further in this paper.

\medskip

\noindent\textbf{Invariant conservative parameterization using inverse classification.} We now focus on the problem of finding elements of the class of shallow-water equations~\eqref{eq:ShallowWaterClosedGeneralForm} with $f$ given by~\eqref{eq:DissipationShallowWater} and $g=0$ that are invariant under a certain subgroup $G^1$ of the maximal Lie invariance (pseudo)group $G$ of the shallow-water equations. That is, we use the inverse symmetry classification strategy to find models from this class. As a symmetry subgroup $G^1$, we single out the four-parameter subgroup of $G$ that is generated by the four-dimensional Lie subalgebra $\mathfrak g^1$ of $\mathfrak g$ with the basis elements
\[
 \p_t,\quad \p_x,\quad t\p_x+\p_u,\quad t\p_t +(1+d)x\p_x + du\p_u+2dh\p_h,
\]
for an arbitrary constant $d\in\mathbb{R}$. The reason for choosing this particular subalgebra is that for physical arguments we require our dissipation scheme to be invariant under the Galilean group (generated by the first three elements of $\mathfrak g^1$) and to have a scaling symmetry. We only require invariance under a single scaling instead of the two scalings admitted by the original shallow-water equations~\eqref{eq:ShallowWater} as adding dissipation to a hydrodynamical system usually breaks one scaling symmetry. For example, the shallow-water system with classical linear dissipation, $f=u_{xx}$ has the above scaling symmetry provided $d=-1/2$.

As discussed in Section~\ref{sec:ShallowWaterInvariantConservative} one can use the replacement theorem to construct parameterization schemes that are both conservative and invariant. To this end, we determine the second order differential invariants of the Lie algebra $\mathfrak g^1$. They can be found by using infinitesimal techniques through prolongation of $\mathfrak g^1$ to the action on second derivatives of $u$ and $h$, and then invoking the infinitesimal invariance criterion~\cite{blum89Ay,blum10Ay,blum74Ay,olve86Ay,ovsi82Ay}. These invariants can also be found using the moving frame method~\cite{cheh08Ay,fels98Ay,fels99Ay}. In particular, on the space of equation variables for the class~\eqref{eq:ShallowWaterClosedGeneralForm} with $f$ satisfying~\eqref{eq:DissipationShallowWater} and $g=0$, which is the subspace of the second jet space $J^2$ with coordinates $(t,x,u,h,u_t,u_x,h_t,h_x,u_{xx})$, there are five elementary invariants given by
\begin{align}
\begin{split}
 &I_1=hu_x^{2d},\quad I_2=h_xu_x^{d-1},\quad I_3=u_{xx}u_x^{-(2+d)},\\
 &I_4=u_x^{d-1}(u_t+uu_x),\quad I_5=u_x^{2d-1}(h_t+uh_x).
\end{split}
\end{align}
The invariant representation of system~\eqref{eq:ShallowWater} is therefore
\[
 I_4+I_2=u_x^{d-1}\Delta_1,\quad I_5+I_1=u_x^{2d-1}\Delta_2
\]
from which it follows that the multipliers $\Gamma^1_1$, $\Gamma^2_1$, $\Gamma^1_2$ and $\Gamma^2_2$ in system~\eqref{eq:InvariantRepresentationDifferentialEquation} are $\Gamma^1_1=u_x^{d-1}$, $\Gamma^2_1=0$, $\Gamma^1_2=0$ and $\Gamma^2_2=u_x^{2d-1}$.

The system of shallow-water equations with the dissipation scheme~\eqref{eq:DissipationShallowWater} is invariant under the group generated by the elements of~$\mathfrak g^1$ provided that
\begin{equation}\label{eq:ConditionForInvariantDissipation}
 \Gamma^1_1f=\tilde f(I_1,I_2,I_3)
\end{equation}
holds for some function $\tilde f$ that can depend at most on the invariants $I_1$, $I_2$ and $I_3$ since none of the functions $f^1$, $f^2$ and $f^3$ in~\eqref{eq:DissipationShallowWater} depends on $u_t$ or $h_t$.

We now determine some dissipation schemes that fulfill the above requirement. For the sake of simplicity, we assume that $f^2=f^3=0$, since we are mainly interested in the second-order term proportional to $u_{xx}$ and thus in finding functions $f^1$ that lead to invariant and conservative diffusion schemes. As none of the invariants $I_1$, $I_2$ and $I_3$ depends on $x$ we have that $f^1_x=0$. The function $\tilde f$ should depend linearly on $I_3$ to match with $u_{xx}$ in $f$. Comparing the coefficients of $u_{xx}$ we find that
\[
 f^1=u_x^{-(1+2d)}\alpha(I_1).
\]
Note that $\alpha$ cannot depend on $I_2$, since $f^1$ does not depend on $h_x$. The remaining condition that has to hold is that
\[
 h_x\int f^1_h\ddd u_x = u_x^{1-d}\beta(I_1,I_2).
\]
The function $\beta$ now cannot depend on $I_3$ as there is no $u_{xx}$ term in the above left-hand side. Since the left-hand side is linear in $h_x$, we find that $\beta = I_2\gamma(I_1)$ and thus get
\[
 \int \alpha_{I_1} u_x^{-1}\ddd u_x = \gamma(I_1).
\]
which imposes a relation between the functions $\alpha$ and $\gamma$. This relation is particularly straightforward to evaluate for polynomial functions $\alpha$. To give an example, let us set $\alpha = 2cdI_1^2$, for $c\in\mathbb{R}$. This leads to the shallow-water equations with dissipation in the form
\[
 u_t + uu_x + h_x = \DD_x\left(ch^2(u_x)^{2d}\right),\quad h_t + uh_x+hu_x=0.
\]

The usual linear dissipation $f=\nu u_{xx}$, $\nu\in\mathbb{R}$, falls into this class when putting $d=-1/2$ and using $\alpha(I_1)=\nu$.

\section{Conclusion}\label{sec:ConclusionConservativeParameterizations}

In this paper we have studied the problem of finding physical parameterization schemes that lead to closed systems of averaged differential equations which possess nontrivial local conservation laws. A main motivation for our work is that by its formulation, one cannot expected to find an exact solution to the parameterization problem. The determination of the entire subgrid-scale structure of a real-world process when one has at their disposal only the grid-scale information is not feasible for nontrivial physical problems. Hence, any auxiliary information that can be used to limit the possible form of a parameterization scheme by imposing some physically and geometrically relevant structural constraints is highly useful. The preservation of symmetries and conservation laws can serve as such relevant constraints since these two properties are closely linked with the physics encoded in a system of differential equations.

A systematic toolbox of methods that allows one to systematically find parameterization schemes with symmetry properties using group classification techniques was formulated in~\cite{bihl11Fy,popo10Cy}. As far as we know, this paper is the first to use the analog toolbox for finding parameterization schemes preserving conservation laws.

The results of this paper illustrate that the requirement of preserving a particular set of conservation laws when constructing physical parameterization schemes can lead to rather specific forms for these schemes. This is in striking contrast to the case of invariant parameterization schemes, since here there are in general an infinite number of possibilities to construct a subgrid-scale closure possessing a prescribed maximal Lie invariance group. Moreover, as to be expected, the more conservation laws one aims to conserve when constructing a subgrid-scale closure or any other additional model for a system of differential equations, the less freedom one has to adjust the closure by including other desirable properties. Thus, the requirement of preserving certain conservation laws (and symmetries) can lead to rather specific parameterization ansatzes which in consequence could potentially simplify the construction and testing procedures for subgrid-scale closure models.

In the present paper, the primary focus of the presentation was to give a careful exposition of the different ideologies for finding conservative closure schemes. The system of one-dimensional shallow-water equations served as a proof-of-the-concept example but did not reveal new physical insights. More realistic examples of conservative parameterization schemes for the governing equations of hydro-thermodynamics will be presented elsewhere.

\section*{Acknowledgements}

For deriving the determining equations of characteristics of conservation laws we have used the package \texttt{GEM} by Alexei Cheviakov~\cite{chev07Ay}. This research was supported by the Austrian Science Fund (FWF), project J3182--N13 (AB) and the National Sciences and Engineering Research Council of Canada (GB).

{\footnotesize\setlength{\itemsep}{0ex}

}

\end{document}